\documentclass[fleqn,10pt]{wlscirep}
\usepackage[utf8]{inputenc}
\usepackage[T1]{fontenc}
\usepackage{tcolorbox}
\usepackage{siunitx}
\usepackage{csquotes}
\definecolor{mBlue}{HTML}{407193} 
\definecolor{mOrange}{HTML}{ca8243} 
\title{Meta-learning ecological priors from large language models explains human learning and decision making}

\author[1,2, 3, 4, *]{Akshay K. Jagadish}
\author[1]{Mirko Thalmann}
\author[1]{Julian Coda-Forno}
\author[1,+]{Marcel Binz}
\author[1,+]{Eric Schulz}
\affil[1]{Institute for Human-Centered AI, Helmholtz Computational Health Center, Munich, Germany}
\affil[2]{Computational Principles of Intelligence, Max Planck Institute for Biological Cybernetics, Tübingen, Germany}
\affil[3]{Eberhard Karls University of Tübingen, Tübingen, Germany}
\affil[4]{Princeton AI Lab, Princeton University, Princeton, USA}
\affil[*]{corresponding author: akshaykjagadish@gmail.com}
\affil[+]{these authors contributed equally to this work}


\begin{abstract}
Human cognition is profoundly shaped by the environments in which it unfolds. Yet, it remains an open question whether learning and decision making can be explained as a principled adaptation to the statistical structure of real-world tasks. We introduce ecologically rational analysis, a computational framework that unifies the normative foundations of rational analysis with ecological grounding. Leveraging large language models to generate ecologically valid cognitive tasks at scale, and using meta-learning to derive rational models optimized for these environments, we develop a new class of learning algorithms: Ecologically Rational Meta-learned Inference (ERMI). ERMI internalizes the statistical regularities of naturalistic problem spaces and adapts flexibly to novel situations, without requiring hand-crafted heuristics or explicit parameter updates. We show that ERMI captures human behavior across 15 experiments spanning function learning, category learning, and decision making, outperforming several established cognitive models in trial-by-trial prediction. Our results suggest that much of human cognition may reflect adaptive alignment to the ecological structure of the problems we encounter in everyday life.
\end{abstract}
\begin{document}

\flushbottom
\maketitle

\thispagestyle{empty}

\section*{Significance Statement}
Humans are remarkably adaptive, making good decisions in complex, uncertain environments. But where do these abilities come from and how can we model them? This work introduces a new approach that combines insights from psychology and machine learning to explain human cognition as an adaptation to ecological environments. By using large language models to generate realistic problems and training neural networks to solve them, we show that simple general-purpose systems can mirror how people learn, categorize, and decide. Our results suggest that much of human learning and decision making may be explained by attunement to the structure of the world around us.

\section*{Introduction}
It is a truth universally acknowledged that a mind in search of a decision is influenced by its environment. Charles Darwin \cite{darwin1859origin} showed that species are adapted to their environmental niche to survive. Egon Brunswik \cite{brunswik1956perception} proposed that people carefully interpret the signals in their surroundings to make judicious decisions. Herbert Simon \cite{simon1956rational} emphasized that human behavior is the result of the interplay between limited cognitive resources and the structure of the environment. Gerd Gigerenzer \cite{gigerenzer1999simple} furthered this notion by introducing the concept of ecological rationality, proposing that minds adapt to their environments by relying on simple context-specific strategies. Yet it remains unclear how attuned human learning and decision making are to the statistical structure of ecologically valid environments.

Two prominent frameworks have sought to address this question through computational modeling: rational analysis \cite{anderson1990adaptive} and ecological rationality \cite{goldstein2002models}. While rational analysis seeks optimal strategies within formal models of the environment, ecological rationality emphasizes heuristics tuned to the structure of real-world tasks. Although rational analysis offers a principled way to derive an adaptive strategy, it requires defining a formal model of the environment. This requirement limits its applicability to relatively simple environments. Ecological rationality, on the contrary, offers a flexible way to model real-world behavior, but it relies on the researcher to hand-design suitable heuristics. This reliance makes it challenging to extend the framework to new domains where effective heuristics have yet to be discerned.

We introduce ecologically rational analysis, a framework that synthesizes the strengths of rational analysis and ecological rationality. This framework enables the automated derivation of computational models that implement approximately optimal strategies directly adapted to the statistical structure of natural environments. These models can subsequently be interrogated -- through classical psychological experiments, for example -- to elucidate how and which environmental properties give rise to human behavior. 

To develop this framework, we draw on two recent advances in machine learning: large language models (LLMs) and in-context learning \cite{brown2020language}. LLMs are generative models trained on internet-scale corpora, capable of capturing the statistical regularities that characterize real-world tasks and domains \cite{Borisov2022-sr, zhu2024eliciting}. We harness this capacity to generate ecologically valid learning environments: problems that approximate the kinds of structure humans are likely to encounter in everyday life \cite{jagadish2024human,marewski2009good}. In-context learning refers to the ability of neural networks to learn from examples presented within a sequence, adapting their behavior purely through internal activations, without any parameter updates \cite{brown2020language}. When derived via meta-learning, in-context learning has been shown to approximate Bayes-optimal inference conditioned on the statistics of its training distribution \cite{ortega2019meta, binz2023meta}.

By meta-learning on tasks generated by LLMs, we develop models that internalize the ecological priors inherent in these environments. We term this class of models Ecologically Rational Meta-learned Inference (ERMI): a family of in-context learners that flexibly adapt to the statistical structure of naturalistic problems. We find that ERMI robustly captures human behavior across 15 experiments encompassing three core domains of cognition: function learning, category learning, and decision making. Beyond accounting for hallmark behavioral signatures within each domain, ERMI yields superior trial-by-trial predictions of human choices relative to a diverse array of established cognitive models. Collectively, these findings suggest that adaptive alignment with environmental statistics is sufficient to account for a broad spectrum of human learning and decision making behavior.

\begin{figure*}
\centering
\includegraphics[width=1\linewidth]{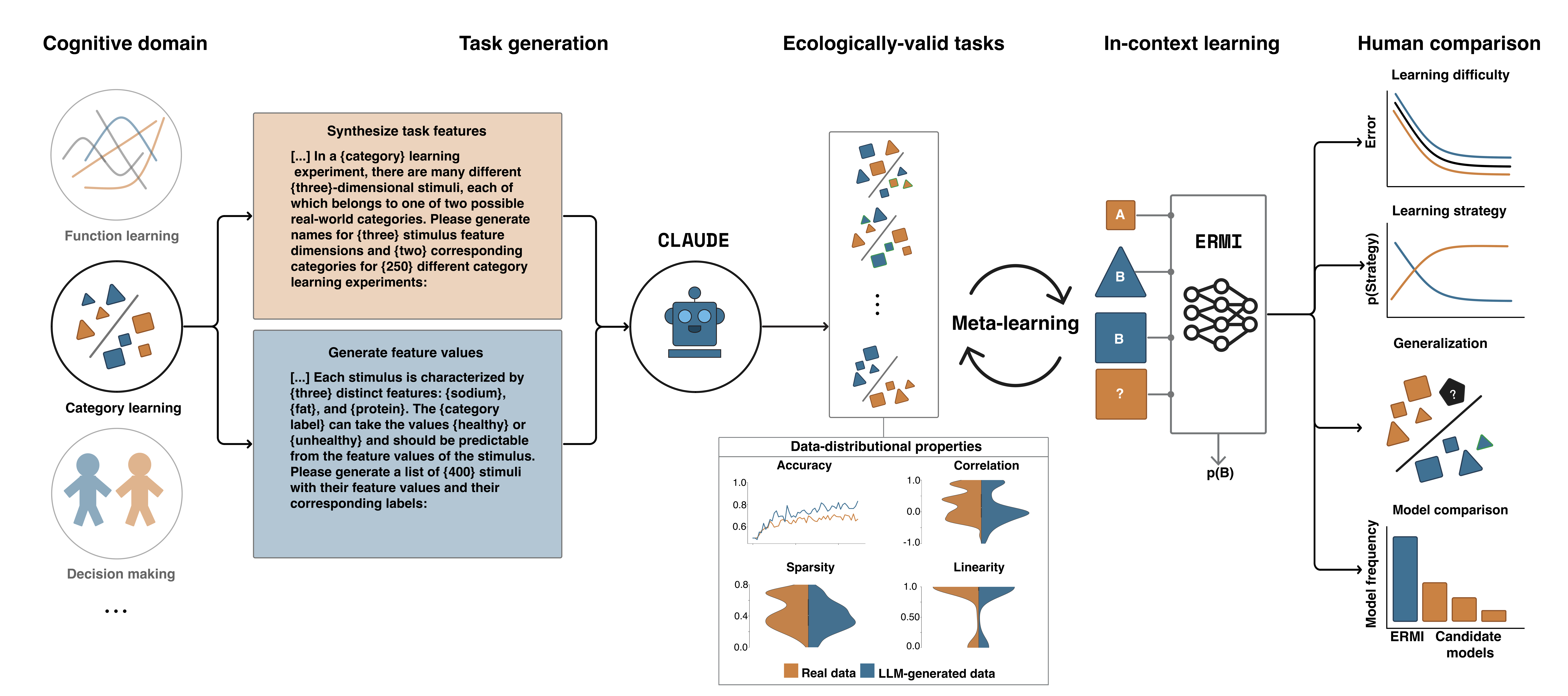}
\caption{\textbf{Schematic of Ecologically Rational Meta-learned Inference}: 
Ecologically Rational Meta-learned Inference (ERMI) is domain-agnostic and can be applied to any cognitive domain. Let us consider category learning as the domain of interest for this illustration. The first step in deriving ERMI is to use a LLM (e.g., \textsc{Claude-v2}) to generate ecologically valid tasks. Task generation from an LLM proceeds in two stages: first, the LLM synthesizes plausible task features (e.g., predicting whether a food item is healthy or unhealthy based on sodium, fat, and protein content); second, it generates corresponding input-target pairs consistent with these features \cite{jagadish2024human}. 
Once a sufficient number of category learning tasks are generated, we analyze their distributional properties, such as classification accuracy, input feature correlation, sparsity in predictive features, and linearity of category structures, and compare them to real-world datasets (e.g., OpenML-CC18\cite{oml-benchmarking-suites}) to verify their ecological validity.
We then derive computational models that internalize these ecological priors by training a neural network (e.g., transformer\cite{vaswani2017attention}) on the LLM-generated tasks using meta-learning. This yields a family of in-context learners, termed Ecologically Rational Meta-learned Inference (ERMI), which flexibly adapt to the statistical structure of naturalistic problems. In category learning, the resulting models are evaluated against human behavior across four key dimensions: learning difficulty, learning strategy, generalization, and quantitative fit to human behavior through model comparison.
}
\label{fig:overview}
\end{figure*}

\section*{Results}

In what follows, we describe ERMI and demonstrate how it can be used to model human learning and decision making across diverse cognitive domains; see Figure \ref{fig:overview} for an overview. 

The core idea behind ERMI is that adaptive behavior reflects the internalization of ecological priors. To capture these priors, ERMI uses LLMs as generative engines to construct ecologically valid tasks (see scalable generation of cognitive tasks from LLMs in Methods). 
This generation process involves two stages: first, the LLM proposes plausible task features (e.g., predicting weight from calories consumed); second, it generates corresponding input-target pairs for the given task features (e.g., specific calorie and weight values) \cite{jagadish2024human}. Importantly, the generated targets correspond to ground truth values and \emph{not} human predictions. 
By querying LLMs, we synthesize a rich and diverse set of cognitive tasks that approximate the distribution of problems found in natural settings.

ERMI then uses meta-learning \cite{hochreiter2001learning, wang2016learning, lake2023human} to derive computational models adapted to the LLM-generated tasks (see Methods). The resulting models implement approximately Bayes-optimal policies that adapt in-context -- modifying internal activations rather than parameters -- to the structure of encountered problems \cite{ortega2019meta, binz2023meta}. This approach allows us to systematically test whether optimal alignment with ecological statistics is sufficient to account for hallmark patterns of human learning and decision making.

In a series of experiments, we demonstrate how ERMI captures human behavior across three core domains of cognition: function learning, category learning, and decision making. 
For each domain, we show that it replicates core behavioral signatures and provides better trial-by-trial predictions of human choices compared to established cognitive models.

\subsection*{Function learning}

Psychologists have been interested in understanding how people learn the functions underlying the association between an input and a target since the 1960s \cite{carroll1963functional}. Much of these studies have focused on mapping a single-dimensional input to a response, called single-cue function learning \cite{koh1991function,brehmer1974hypothesis}, which is also the focus of this work. 

In these tasks, participants observe input-output pairs, typically receiving feedback on the true response after each prediction. The underlying function is unknown and must be inferred through trial and error. Once trained, participants are tested on previously unobserved inputs within the range of their prior experience (interpolation) or outside that range (extrapolation). Previous work has revealed several hallmark findings: people interpolate more accurately than they extrapolate, sometimes performing as well on novel interpolated inputs as on the training set itself \cite{delosh1997extrapolation}; and they exhibit systematic biases when generalizing, favoring linear functions with positive slopes and minimal offsets \cite{brehmer1974hypothesis, schulz2017compositional, schulz2015assessing}.

\begin{figure*}
\centering
\includegraphics[width=1\linewidth]{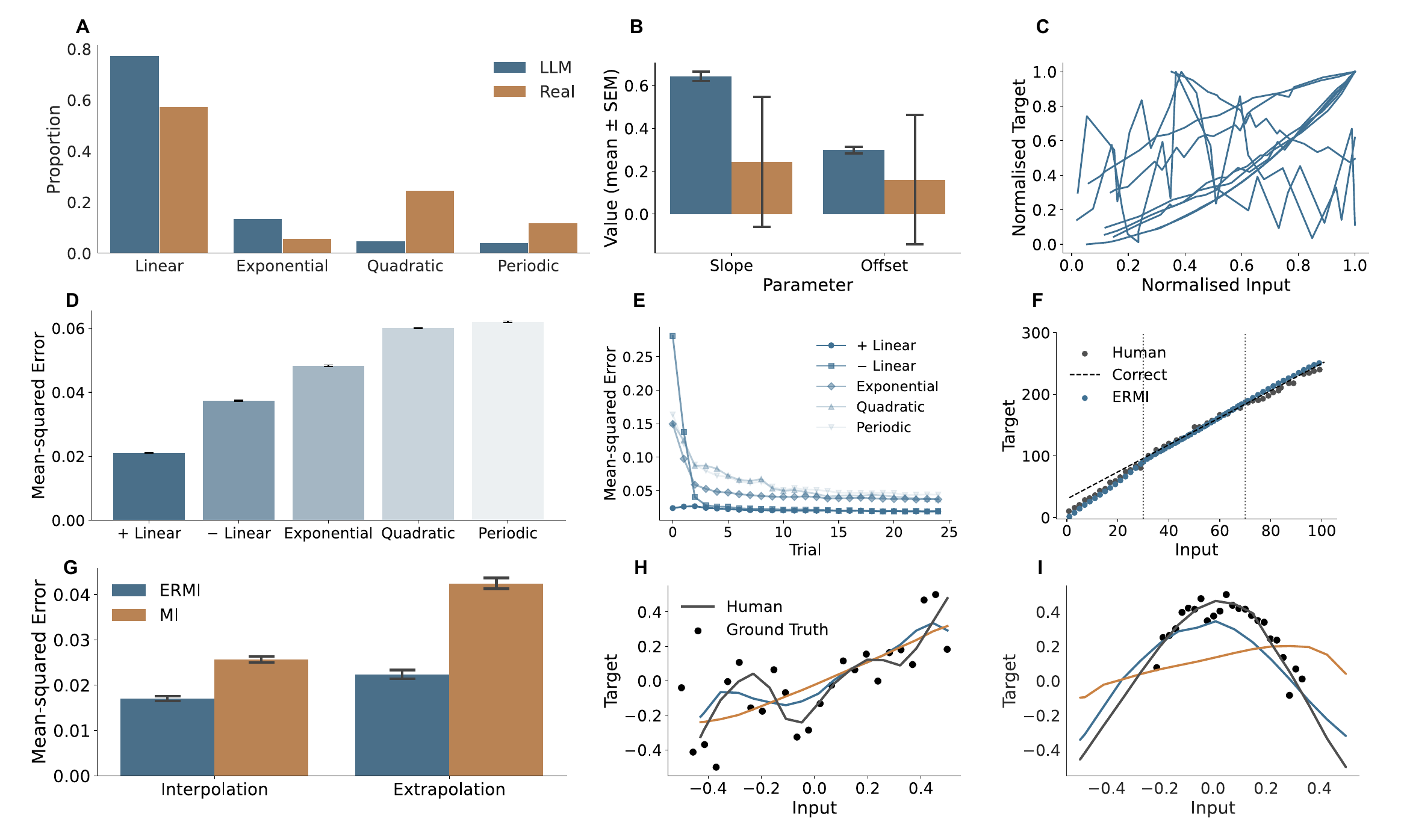}
\caption{\textbf{Function Learning}:\textbf{A} Proportions of different function types in real-world datasets \cite{lichtenberg2017simple} and LLM-generated datasets. \textbf{B} Parameters for slope and offset for linear functions fitted to both datasets. \textbf{C} Example functions sampled from the LLM-generated datasets. \textbf{D} Mean-squared error (MSE) of ERMI when simulated on five function types, namely, positively-sloped linear, negatively-sloped linear, exponential, quadratic, and periodic functions (mean over all runs and trials). \textbf{E} MSE of ERMI for the five function types mentioned above unrolled over trials (mean over all runs). \textbf{F} Simulations of ERMI on linear function from \cite{kwantes2006extrapolation} along with human predictions extracted from the original plot (gray lines mark interpolation range between 30 to 70 and extrapolation region between 0 to 30 and between 70 to 100). \textbf{G} MSE during interpolation (Experiment 2 of original study) and extrapolation (Experiment 4) for ERMI and Meta-learned inference (MI) with hand crafted prior when simulated on tasks from \cite{little2024function}. \textbf{H} Representative example for interpolation. \textbf{I} Representative example for extrapolation.
}
\label{fig:function_learning}
\end{figure*}

ERMI provides a framework for probing the origins of these behavioral patterns. Following the procedure outlined previously, we first construct a dataset of about \num{10,000} one-dimensional function learning tasks designed to reflect the diversity of functional relationships found in natural environments (see Figure \ref{fig:function_learning}C for examples). We analyze the statistical properties of the LLM-generated tasks and compare the tasks with real-world regression problems \cite{lichtenberg2017simple}. We found that the generated functions comprised approximately 75\% linear, 12\% exponential, 7\% quadratic, and 6\% periodic relationships (see Figure \ref{fig:function_learning}A), a distribution that mirrors both environmental regularities and human difficulty rankings in function learning \cite{busemeyer2013learning}. A model-based analysis of linear fits revealed a pronounced bias toward positive slopes with near-zero offsets (see Figure \ref{fig:function_learning}B), consistent with known human biases in extrapolation \cite{kwantes2006extrapolation, kalish2004population}.

Then, we asked to what extent ERMI can replicate the characteristic patterns of human function learning?  
Drawing on prior work \cite{busemeyer2013learning}, we focused on five well-established findings in function learning:
% (1) continuous mappings are learned more easily than categorical ones \cite{carroll1963functional}; 
(i) linearly increasing functions are learned more readily than decreasing functions \cite{brehmer1974hypotheses, brehmer1973single, brehmer1974effects}; (ii) linearly increasing functions are also learned faster than nonlinearly increasing functions \cite{delosh1997extrapolation, byun1995interaction}; (iii) monotonic functions are learned more effectively than non-monotonic ones \cite{carroll1963functional, byun1995interaction, brehmer1974hypotheses}; (iv) cyclic functions are more challenging than non-cyclic functions \cite{byun1995interaction}; and (v) generalization is more accurate in interpolation than in extrapolation \cite{carroll1963functional, delosh1997extrapolation, mcdaniel2005conceptual}.

To test whether ERMI reproduces these patterns, we sampled functions from each class and assessed the model’s learning dynamics. We measured learning speed and accuracy using the rate of change and mean-squared error (MSE) across trials (see Methods for details). Strikingly, ERMI mirrored human behavior across all five phenomena: (i) it learned positive linear functions (M$_{\text{MSE}}$=0.5260, SEM=0.0043, t=-70.1587, p$<$0.001) better than negative linear functions (M$_{\text{MSE}}$=0.9339, SEM=0.0039); (ii) it grasps linear functions with positive slopes more rapidly, reaching minimum MSE in fewer trials (M$_{\text{trial}}$=13.29, SEM=0.051) compared those with negative slopes (M$_{\text{trial}}$=14.22, SEM=0.048; t=-13.38, p$<$0.01); (iii) it mastered monotonic functions (M$_{\text{MSE}}$=0.8895, SEM=0.0028, t=-145.5417, p$<$0.01) more accurately than non-monotonic ones (M$_{\text{MSE}}$=1.5255, SEM=0.0032); (iv) it learned non-cyclic functions (M$_{\text{MSE}}$=1.0423, SEM=0.0025, t=-89.6918, p$<$0.01) more readily than cyclic ones (M$_{\text{MSE}}$=1.5508, SEM=0.0054, t=-89.6918, p$<$0.01); and (v) it achieved better prediction performance during interpolation (MSE=0.0017\cite{kwantes2006extrapolation}) than during extrapolation (MSE=0.0022); see Figure \ref{fig:function_learning}D-E.

% Experiment 4
A well-established finding in the function learning literature is that humans tend to underestimate functional relationships during extrapolation, particularly for linear functions, with a characteristic bias toward zero offset \cite{kwantes2006extrapolation}. To assess whether ERMI exhibits a similar pattern, we conditioned the model on input-target pairs sampled from a linear function, following the procedure of Kwantes and Neal \cite{kwantes2006extrapolation} (see human studies in SI for additional details), and simulated its predictions for input values outside the training range. We compared ERMI's responses to human data in both interpolation and extrapolation regimes. For human responses, we drew on data from Experiment 2 of Kwantes and Neal \cite{kwantes2006extrapolation}, in which participants directly estimated numerical values for a given input -- an evaluation method that closely parallels our procedure for ERMI. We found that, like humans, ERMI systematically underestimated responses in the extrapolation range, with a stronger bias in the lower region (MSE=$0.0043$; 0-30 on the x-axis, which is marked using grey lines in the Figure \ref{fig:function_learning}F) than in the upper region (MSE=$0.0003$; 70-100 on the x-axis). It showed a bias towards zero offset comparable to that observed in human participants (see Figure \ref{fig:function_learning}F). In addition, ERMI's predictions agreed better (MSE=$0.0002$) with human responses than a meta-learned (MI; MSE=$0.00054$) with hand-crafted prior used by Lucas and colleagues \cite{lucas2015rational}; see Methods for details.
% ; see the Inset of Figure \ref{fig:function_learning}F.

% Experiment 2 and 3
Beyond qualitative signatures, a critical test of any model is whether it can capture the fine-grained structure of human behavior at the trial level. To this end, we evaluated ERMI on the function estimation task introduced by Little and colleagues \cite{little2024function}. In this task, participants viewed 24 scatter plots, each depicting data from a different fictional scientific experiment, and were asked to draw what they believed to be the true underlying causal function. The scatter plots were generated from linear, quadratic, or cubic polynomial functions with added Gaussian noise. Crucially, the scatter plots were presented in two distinct formats. In the zoomed-in condition, data points filled the entire plotting area, enabling assessment of interpolation. In the zoomed-out condition, data points were centrally located and occupied only 40\% of the plot area, encouraging participants to extrapolate beyond the range of the observed data; see human studies in the SI for additional details. 

We conditioned ERMI on the same training data shown to participants and generated predictions for the identical input values. To quantify model fit, we computed the MSE between ERMI’s predictions and human responses across all test inputs. As shown in Figure~\ref{fig:function_learning}G, ERMI provided a closer fit to human judgments than MI in both interpolation and extrapolation conditions. Specifically, ERMI achieved a lower MSE (M$_{\text{MSE}}$=0.0171, SEM=0.0014, t = -9.9944, p $<$ 0.01) compared to MI (M$_{\text{MSE}}$=0.0256, SEM=0.0020) during interpolation, and likewise outperformed MI during extrapolation (ERMI: M$_{\text{MSE}}$=0.0223, SEM=0.0018; MI: M$_{\text{MSE}}$=0.0424, SEM=0.0034, t = -13.1479, p $<$ 0.01). For illustrative examples, Figure~\ref{fig:function_learning}H-I shows ERMI’s predictions, alongside those of MI, for interpolation and extrapolation condition from a representative participant; see Figure S2 in the SI for additional examples.
Together, these findings suggest that a rational model attuned to the statistics of ecologically valid function learning problems is sufficient to capture much of human function learning.

\subsection*{Category learning}

To examine whether these findings generalize, we turned to a second domain that has been widely studied in cognitive psychology, namely category learning \cite{ashby_human_2005}. In a typical category learning task, participants are presented with inputs sequentially and must assign each one to a set of known categories. After each trial, they receive feedback that indicates whether their classification was correct. The underlying rule that governs category membership -- referred to as the category structure -- is hidden from participants and must be inferred through trial and error. During the test phase, participants classify both previously seen (training) and novel (transfer) input without receiving feedback. This design allows for simultaneous assessment of learning performance on familiar examples and generalization to new instances.

It has been observed that humans find learning certain category structures more difficult than others \cite{Shepard1961-yu}. Furthermore, the categorization strategy they use changes during the course of the experiment from an exemplar-based strategy to a prototype-based strategy \cite{smith1998prototypes}. In addition, the way they generalize to unseen inputs is systematic, following a rule-plus-exception-based model \cite{Johansen2002-xe}. 

\begin{figure*}
\centering
\includegraphics[scale=0.40]{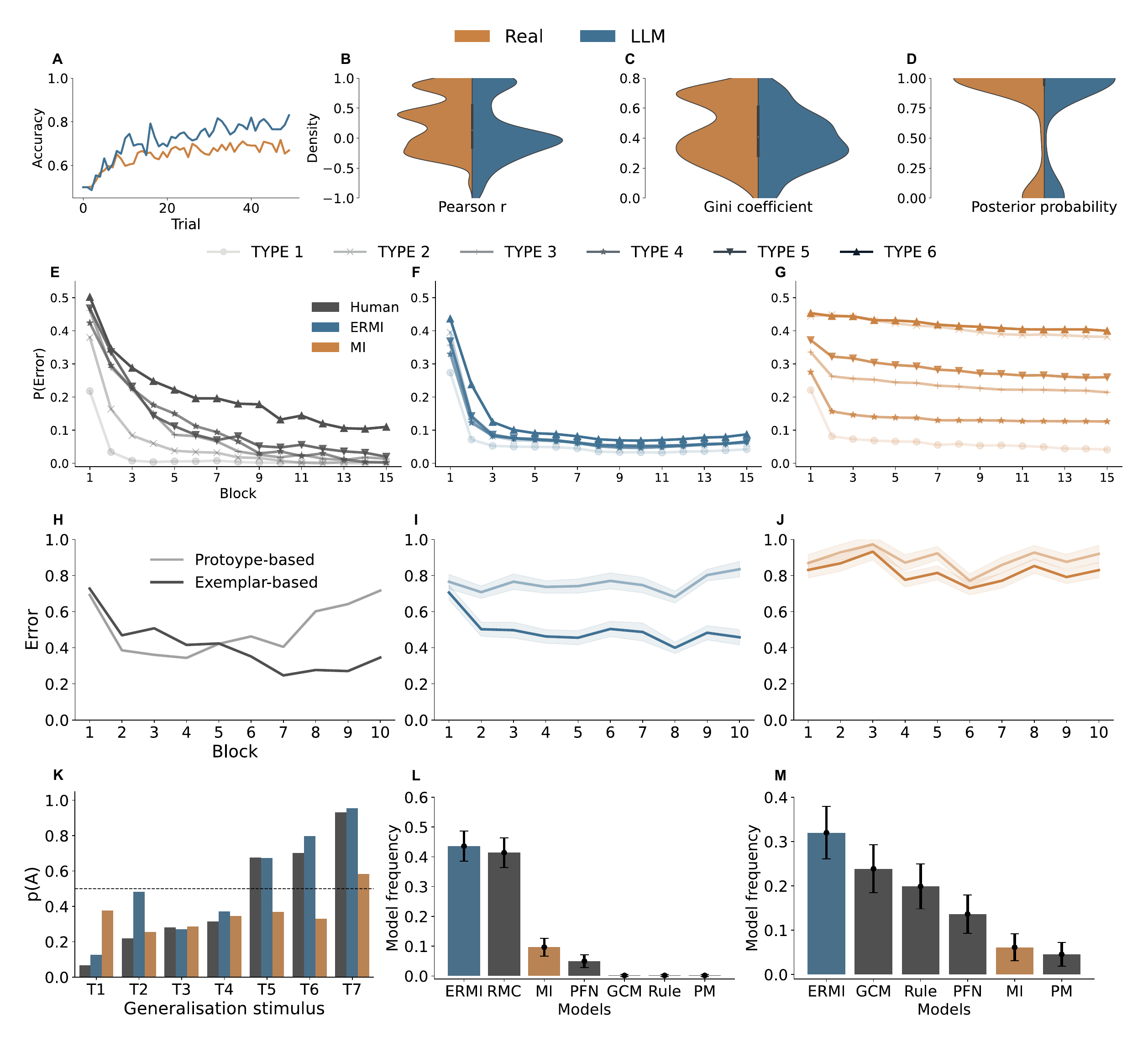}
\caption{\textbf{Category learning}:\textbf{A} Mean task performance of a logistic regression model over trials for real-world classification tasks \cite{oml-benchmarking-suites} in orange and LLM-generated tasks in blue. \textbf{B} Density plot of Pearson's correlation coefficients between feature dimensions. \textbf{C} Gini coefficients over logistic regression weights, which provides a measure of sparsity (high values indicates greater sparsity). \textbf{D} Posterior probability measuring the linearity of category learning tasks. \textbf{E–G} Average error probabilities for each task \textsc{type} across 16-trial blocks for \textbf{E} humans (in gray), \textbf{F} ERMI (in blue), and \textbf{G} MI (in orange). Human data in \textbf{E} reproduced from Table 1 in \cite{Nosofsky1994-hw}. ERMI and MI were simulated on \textsc{type 1–6} tasks for 50 runs using the inverse temperature ($\beta$) that minimized mean-squared error with respect to human data: $\beta=0.4$ for ERMI and $\beta=0.9$ for MI.
\textbf{H–J} Average error of exemplar- and prototype-based models fitted to \textbf{H} human choices, \textbf{I} simulated choices from ERMI, and \textbf{J} simulated choices from MI across 56-trial blocks. Human data in \textbf{H} reproduced from \cite{smith1998prototypes}. ERMI and MI were simulated using inverse temperature values fitted to participants’ choices in \cite{devraj2021dynamics}; ERMI (M$_{\beta}$=$0.09$, SEM=$0.01$) and MI (M$_{\beta}$=$0.17$, SEM=$0.02$). Shaded regions indicate standard error of the mean.
\textbf{K} Average categorization probabilities of transfer inputs T1–T7 for humans (gray), ERMI (blue), and MI (orange). Human data reproduced from \cite{Johansen2002-xe}. ERMI and MI were simulated on the same experiment over 77 runs using the best-fitting inverse temperatures: $\beta=0.9$ for ERMI and $\beta=0.1$ for MI.
\textbf{L} Posterior model frequency of participants’ choices in \cite{Badham2017-hc} across seven computational models. 
\textbf{M} Posterior model frequency of participants’ choices in \cite{devraj2021dynamics} across six computational models.
}
\label{fig:category_learning}
\end{figure*}

%% data stats (not necessary) I could allude to previous work
To investigate the role of ecological adaptation in explaining these findings, we again turn to ecologically rational analysis. 
% We adapt the procedure described earlier to generate category learning problems from an LLM, verify their ecological validity, and then derive ERMI by meta-learning on the LLM-generated problems.
Following a previously established procedure, we generated about \num{10,000} category learning problems and inspected their underlying statistics. Like in function learning, we found that LLM-generated category learning problems capture key statistical properties of real-world classification datasets \cite{oml-benchmarking-suites}. Specifically, we observe that (i) the generated category learning problems are noisy, yet classifiable, like real-world classification problems (Figure \ref{fig:category_learning}A); (ii) they contain inputs whose features exhibit a full range of correlations, from non-existent or low values to nearly complete correlation, as in real-world tasks (Figure \ref{fig:category_learning}B); (iii) only a few feature dimensions within each task substantially contribute to classification, indicating sparsity in the feature space commonplace in real-world tasks (Figure \ref{fig:category_learning}B; larger Gini coefficient values indicate higher sparsity); and (iv) the category structure observed in the generated classification problems is predominantly linear akin to real-world tasks (Figure \ref{fig:category_learning}B; higher values indicate more linearity).
These findings confirm the ecological validity of LLM-generated category learning tasks. 

%% results from ERMI
Moving to the next step, we derived ERMI by meta-learning on these category learning problems and evaluated how well it can capture various aspects of human category learning.
To explain human learning difficulties during category learning using ERMI, we consider the study by Shepard et al. \cite{Shepard1961-yu}. In this study, the authors considered six different category structures (labeled \textsc{type 1} to \textsc{type 6}). In \textsc{type 1} problems, all items in a category share one particular feature value (e.g., they are all black), \textsc{type 2} problems are defined by a combination of two feature values (i.e., XOR problems), \textsc{type 3-5} problems combine a rule with exceptions, and \textsc{type 6} problems require the memorization of individual items as rules and the similarity to other items is not informative; see human studies in the SI for details.
% the first structure assigning a input to a category based on a single feature and the sixth structure. 
The task difficulty of the six problem types increases from 1 (M$_{\text{p(Error)}}$ = 0.0201) to 6 (M$_\text{p(Error)}$ = 0.2048), as shown in Figure \ref{fig:category_learning}E. The error rates for \textsc{type 2-5} problems fall between those of \textsc{type 1} and \textsc{type 6}; see Table S3 in SI for details. 
 ERMI -- when simulated on tasks from the Shepard et al. study \cite{Shepard1961-yu} -- displayed learning curves that are difficulty dependent and follow the same ordering as people's; see Figure \ref{fig:category_learning}F. Quantitatively, ERMI ($\mathrm{MSE} = 0.03$) captured human learning difficulties better than meta-learned inference with a hand-crafted prior (MI; $\mathrm{MSE} = 0.26$); see Methods for details.
% This is a key finding that provides the need for ERMI. RMC is trained on samples from the true generative model of the task. This results indicates that LLM-generated category problems capture additional statistics --beyond those used to create the tasks-- that are necessary to explain human behavior. 
% ERMI's ($\mathrm{MSE} = 0.03$)) learning curves was much similar to humans than that of RMC ($\mathrm{MSE} = YY$). 
% The learning curves of other baseline models matched humans to a lesser extent; see Table 1 in the SI. 

We then investigated whether ERMI also captures human trial-by-trial choices during category learning by considering a replication of the original Shepard's study \cite{Shepard1961-yu} by Badham et al. \cite{Badham2017-hc}; see human studies in the SI for details. 
We performed a Bayesian model comparison between ERMI and five other computational models, which included the rational model of categorization (RMC\cite{Anderson1991-ii}), a prototype model (PM\cite{homa1984role}), an exemplar model (generalized context model (GCM)\cite{nosofsky1986attention}), a rule-based model (Rule\cite{mcdaniel2005conceptual} and a meta-learned inference model with Bayesian logistic regression prior (MI) and Bayesian neural network prior (PFN\cite{Muller2021-ol}); see Methods for details on the baseline models, as well as the model fitting and comparison procedure.
We found that in terms of the posterior model frequency (PMF), which measures how often a model offers the best explanation in the population, ERMI explains human choices more frequently (M$_{\text{PMF}}$=$0.43$, SEM=$0.05$) compared to the other models, with the RMC coming in close second (M$_{\text{PMF}}$=$0.41$, SEM=$0.05$); see Figure \ref{fig:category_learning}L.  

After that, we tested whether ERMI shows a similar shift in its categorization strategy as humans \cite{smith1998prototypes}. In this study, participants classified 14 six-dimensional inputs into two categories. These categories were assigned based on a nonlinear decision rule; see human studies in the SI for details.
The authors then fitted a prototype and an exemplar model to the observed behavior and found that the prototype model better explained the people in the early blocks but in the later blocks, their choices aligned more closely with the exemplar model, as shown in Figure \ref{fig:category_learning}H. When simulated on tasks from the same study, we found ERMI's strategy to be indistinguishable between prototype-based and exemplar-based in the beginning of the experiment, but with experience, it became increasingly more exemplar-based as observed in humans (see Figure \ref{fig:category_learning}I). In contrast, MI does not display such a transition in strategy, as shown in Figure \ref{fig:category_learning}J. Furthermore, we compared ERMI with other competing models in the prediction of human choices at the trial level, for which we used human data from a replication of the original study by Devraj et al. \cite{devraj2021dynamics}. As shown in Figure \ref{fig:category_learning}M ERMI (M$_{\text{PMF}}$=$0.32$, SEM=$0.06$) predicted human choices the most frequently, followed by the GCM (M$_{\text{PMF}}$=$0.24$, SEM=$0.05$) and the rule-based model (M$_{\text{PMF}}$=$0.20$, SEM=$0.05$).

Finally, we examined whether ERMI displays the same generalization patterns as people when they observe inputs not part of the training phase \cite{Johansen2002-xe}. 
In the training phase of this study, participants performed binary classification of nine four-dimensional inputs. Subsequently, in the test phase, they were probed on seven transfer inputs (labeled T1-T7; see Methods for their encodings) for which they did not receive any feedback; see human studies in the SI for details. The latter was intended to examine how they would generalize the learned category structure to unseen inputs; see Method for details. 
Figure \ref{fig:category_learning}K shows the proportion of responses in which the participants assigned category A to the seven transfer inputs (in gray). It can be seen that participants assigned the transfer inputs T5, T6, and T7 mainly to category A and the inputs T1, T2, T3, and T4 mainly to category B. 
% We provide further details about the paradigm in Appendix \ref{app:experiment3}.
ERMI -- when evaluated on the same task -- generalizes to unseen inputs in a human-like way by classifying the inputs T1, T3, and T4 more often as category B and the inputs T5, T6, and T7 more often as category A; see Figure \ref{fig:category_learning}I (in blue). 
The only deviation from human-like generalization is input T2. Although ERMI classified it as category A at the chance level, the participants predominantly assigned it to B. We speculate that this is because T2 resembles the category prototype along two only dimensions, while other inputs categorized as B matched along three dimensions. 
% unlike other inputs categorized as B that resembles category prototype on three dimensions, T2 matches only on two.
% We speculate that this may be due to input T2 only having two non-zero features unlike all other inputs categorized as B, which contained three non-zero features.
Yet again, MI did not show the same pattern as in humans, both qualitatively (Figure \ref{fig:category_learning}K; in orange) and quantitatively, with the Euclidean distance between the choice probabilities of humans and ERMI ($0.29$) being lower than between humans and MI ($0.67$).

These results indicate that in addition to capturing human function learning, ERMI also captures human category learning.

\subsection*{Decision making}
The question of how people decide between multiple options and how they improve on it with experience has been extensively studied in economics \cite{samuels2012ending}, psychology \cite{gigerenzer1999simple}, and neuroscience \cite{camerer2005neuroeconomics}. Does ERMI also extend to this domain?

For our analyses, we considered the paired comparison task \cite{binz2022heuristics}. Participants in this task decide between two options, each of which is characterized by different feature values.  The feature values are associated with a value on an unobserved criterion and participants have to learn which option has the higher criterion. We consider a sequential variant where participants take decisions one at a time with feedback provided on which option had a higher criterion value after each trial. 

\begin{figure*}
\centering
\includegraphics[scale=0.38]{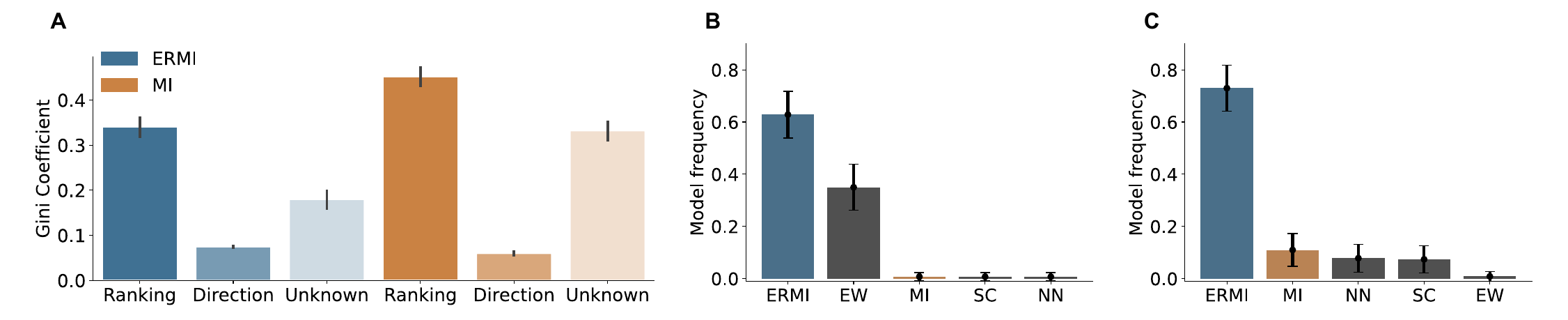}
\caption{\textbf{Decision Making}:  \textbf{A} Mean Gini coefficients computed over weights produced ERMI and MI on the tenth trial of the paired comparison tasks sampled from the generative model used in the Binz and colleagues \cite{binz2022heuristics} for three conditions: ranking, direction and unknown. \textbf{B} Posterior model frequency of participants’ choices from experiment 3b of Binz et al. study \cite{binz2022heuristics}, which uses four attributes for each option. \textbf{C} Posterior model frequency of participants’ choices from experiment 3a of Binz et al. study \cite{binz2022heuristics},  which uses two attributes for each option. }
\label{fig:decision_making}
\end{figure*}

% some findings
% heuristics: one-reason decision making, equal weighted decision making comes up under different conditions
The strategies people use to make decisions in a paired comparison task are widely contested. While economists have taken a rational perspective that suggests that people weigh the different cues appropriately while making decisions \cite{samuels2012ending, geisler1989sequential}, proponents of ecological rationality have argued that people are incapable of such reasoning due to their cognitive constraints \cite{gigerenzer2011heuristic, chater2003simplicity}. 
Instead, they have proposed the view that people rely on simple heuristics, which are short-cut strategies that produce competitive performance despite using only parts of the available information \cite{todd2007environments}. 
% Previous work has demonstrated that people deploy decision making strategies, such as one-reason or equal weighting, in a context-specific way in paired comparison settings \cite{binz2022heuristics, martignon2002fast}.
Recent work  \cite{binz2022heuristics} has shown that people adopt different decision-making heuristics depending on the structure of paired comparison tasks \cite{martignon2002fast}. When participants knew the importance of attributes to the criterion but not the direction of their correlation with it (ranking condition), they used a one-reason decision strategy. When the direction was known but not the ranking (direction condition), they relied on an equal weighting strategy. Finally, when neither ranking nor direction was known (unknown condition), they used a weighted combination of attributes to guide their choices.

To further examine how data distributional properties influence heuristic choice, we adapted the previously described procedure to generate three sets of problems, each containing approximately \num{7,000} tasks, reflecting one of three conditions: ranking, direction, and unknown. This was done using three condition-specific prompts specifying: (i) that attributes be rank-ordered by importance to the target; (ii) that attributes correlate positively with the target; or (iii) no additional information to allow free-form generation. We then verified that the generated tasks matched their intended conditions: tasks in the ranking condition showed more rank-ordered feature weights than those in the unknown condition, and tasks in the direction condition had features more strongly (positively) correlated with the target (see Figure S7 in the SI). After that, we constructed paired comparison trials by randomly sampling two options from each dataset and pitting them against each other. We then derived an ERMI model by meta-learning on LLM-generated tasks for each condition.

The resulting ERMI models were first simulated on decision making problems from Binz and colleagues \cite{binz2022heuristics}. 
To examine the strategy being implemented by ERMI, we computed the Gini coefficient over attribute weights produced by ERMI (see Methods for details).
Higher values for the Gini coefficient indicate more sparse weights, which corresponds to a one-reason decision making strategy, lower values correspond to equally weighted attributes, and values in between correspond to a weighted-additive strategy. As shown in Figure \ref{fig:decision_making}A, we found that ERMI uses the same heuristics that people use in the respective condition. That is, ERMI trained on decision making problems from the ranking condition implements a one-reason decision making strategy (M$_{\text{Gini}}$ = 0.3399, SEM = 0.0631), in the direction condition it uses an equal weighting strategy (M$_{\text{Gini}}$ = 0.0743, SEM = 0.0138), and in the unknown condition it relies on a weighted combination of attributes (M$_{\text{Gini}}$ = 0.1794, SEM = 0.0333). These results are also consistent with the strategies used by meta-learned inference (MI) with a hand-crafted prior for each condition; see Methods for details.

In addition to the simulation study, we evaluated whether ERMI explains human decision making better than competing models by conducting a model comparison on human data from Binz et al. \cite{binz2022heuristics}. 
We considered human data from two experiments, one with options containing two attributes and the other with four, in which participants performed decision-making tasks without receiving any side information (see human studies in SI for details).
Compared to other baseline models -- namely, single-cue decision making (SC), equal-weighted strategy (EW), feedforward neural network (NN), and MI -- ERMI accounts for human responses most frequently in both the two-attribute experiment (M$_{\text{PMF}}$=$0.7299$, SEM=$0.0888$) and the four-attribute experiment (M$_{\text{PMF}}$=$0.6284$, SEM=$0.0897$). These results suggest that ERMI converges to the same decision making strategies that people have been shown to use, with the particular strategy shaped by the statistical structure of decision-making problems.

\section*{Discussion}

In the late 1990s, Gerd Gigerenzer and his colleagues conducted what, in hindsight, stands as one of psychology’s most endearingly simple yet profoundly revealing studies. They asked participants whether they recognized the names of various cities or companies—and found that when people chose the company they recognized between two options, their choices reliably predicted which company had higher stock returns. This surprisingly effective strategy, dubbed the recognition heuristic, is a lexicographic decision rule similar to the ones we modeled that stops at the first discriminating cue—in this case, recognition. But how could such a seemingly naive rule succeed in complex contexts like financial forecasting? Gigerenzer proposed that the frequency with which people encounter names in everyday life --on television, in conversations, or in headlines-- contains statistical signals informative for decision making \cite{goldstein1999recognition}, a hypothesis he substantiated by meticulously counting company name occurrences in major newspapers.

Yet this insight raised a far-reaching question: could one ever scale ecological rationality beyond a single heuristic to explain the full complexity of human learning and decision making? After all, manually tallying newspaper mentions might work for isolated cues, but becomes hopelessly unwieldy when faced with the rich environments people face daily. How could one map the statistical fingerprints of vast environments across countless domains of cognition?

In recent years, an unexpected answer may have emerged: large language models. Indeed, it has been argued, by Alison Gopnik and colleagues \cite{farrell2025large}, that LLMs, trained on the sprawling archive of human culture, can be seen as ``cultural technologies,'' artifacts that distill the collective knowledge of societies. Where earlier researchers scraped headlines to estimate how often a name appeared in people’s environments, LLMs now embed billions of such frequencies and co-occurrences, capturing statistical regularities on a scale previously unimaginable. This extraordinary capacity opens, for the first time, an opportunity to massively scale up ecological rationality. We can use LLMs to generate ecologically grounded tasks that reflect the natural statistics of human environments and test whether human learning and decision-making align with ideal inference under such ecological priors.

In the current work, we introduced ecologically rational analysis, a framework that leverages meta-learning and LLMs to do exactly that, i.e. to extend the logic of ecological rationality beyond individual heuristics and into the broader structure of human cognition. In particular, we developed a new class of models—called ERMI—that allowed us to investigate whether human learning and decision making approximate ``ideal statistical inference under the structure of natural tasks and environments'' \cite{tenenbaum2021homepage}. Across 15 experiments spanning three core domains of human cognition, we found that ERMI can account for a substantial amount of variance in human behavior. Not only did ERMI capture key behavioral signatures in each domain, but it also provided superior trial-level prediction of human choices relative to established cognitive models. Taken together, these findings demonstrate that rational adaptation to ecologically valid task statistics is sufficient to account for much of human cognition.

A key strength of ERMI lies in its ability to derive priors and distill them into computational models without extensive hand-engineering. In contrast, traditional rational analysis requires researchers to manually specify the underlying data-generating distribution. For example, in the rational model of function learning, Lucas et al. \cite{lucas2015rational} assumed a linearity-based prior but acknowledged uncertainty about its alignment with naturalistic environments, noting that “it is not realistic to directly measure the statistical structure of the environment, that is, what functions are truly more or less common” \cite{lucas2015rational}. ERMI circumvents this issue by using LLMs to directly generate tasks with ecological statistics. 
Alternatively, ecological rationality often relies on researchers to manually construct heuristics that are ``applicable to specific decision tasks and in particular domains—different tools for different tasks'' \cite{todd2016building}. By leveraging meta-learning to automatically derive computational models adapted to these ecological statistics, ERMI eliminates the need for hand-designing task-specific heuristics prevalent in ecological rationality framework.

Yet one may ask: Why not directly use an LLM to model human behavior, instead of meta-learning on LLM-generated tasks? We evaluated LLMs as direct behavioral models and found that ERMI consistently outperformed them in explaining human data (see Figure S5 in the SI), highlighting that LLMs do not capture human behavior out-of-the-box. Furthermore, even a strong alignment of LLMs with human behavior would not clarify why they are good models, given that they are trained on vast and opaque datasets that span human conversations, code, and various cultural artifacts that are difficult to analyze. ERMI, on the other hand, leverages LLMs solely as generative sources for ecological tasks used in meta-learning, allowing us to test a specific hypothesis about human cognition. Of course, what features of the environment are required to explain behavior in a particular domain still needs to be determined statistically, which re-introduces a degree of manual analysis that ERMI does not yet fully automate. In that sense, while ERMI reduces the burden of handcrafting heuristics and priors, it does not eliminate the need for scientific judgment in feature selection and interpretation. However, this remaining bottleneck also presents a new opportunity: by systematically varying environmental features across LLM-generated tasks and analyzing their impact on model fit and human behavior, we can begin to reverse-engineer the ecological ingredients that most shape cognition.

Future work should determine which additional components are required to account for human behavior beyond ecological rationality. ERMI offers a flexible foundation for integrating such components. First, we can incorporate participant-specific information into the data generation process, followed by meta-learning on these tailored datasets. This approach would enable personalized ERMI models that capture individual differences, particularly those shaped by environmental and demographic factors unique to each participant. Second, while computational models derived via ERMI currently emphasize adaptation to the environment, they largely ignore the role of cognitive constraints. Incorporating limits on computational complexity --such as attention, working memory, or representational capacity-- could help explain additional variance in human behavior, especially in cases where people systematically deviate from ideal inference \cite{binz2022heuristics, jagadish2023zero, jagadish2025bounded}. Notably, such constraints can either be explicitly modeled within the meta-learning process or may already be implicitly embedded in the training data itself, given that these data are generated by humans who are inherently resource-bounded. Third, ERMI could serve as a starting point for fine-tuning on human choice data, following recent approaches \cite{peterson2021using, ji2023automatic, miller2023cognitive}. This would allow for principled estimation of residual variance in behavior not yet captured by ERMI and help identify the cognitive mechanisms needed to close that gap.

Humans excel at learning and decision making in complex and uncertain environments. Our findings suggest that these cognitive abilities emerge largely through attunement to ecological structures. By harnessing ecologically rational analysis, combining psychology and machine learning, we demonstrate how general-purpose models, trained in realistic ecological tasks, can mimic much of human behavior. Looking ahead, we speculate that scaling up this approach to open-ended, embodied environments \cite{team2023human}, which require processing high-dimensional visual information and executing complex control sequences,  holds promise for expanding ecologically rational analysis. By leveraging multimodal foundation models to generate personalized, ecologically valid tasks and meta-learning for distilling those priors into adaptive computational models, we can systematically quantify how much of human behavior can be explained as an adaptation to previously encountered task structures and environments. If successful, this would significantly broaden the explanatory power of cognitive models, offering nuanced insights into the ecological roots of human cognition.

\section*{Methods}
In this section, we first describe how we generate cognitive tasks at a scale that is sufficient to train in-context learning models from scratch and how we verify their ecological validity. Following this, we discuss how to derive in-context learning models via meta-learning and present other domain-specific cognitive models used as baselines. We then describe (a) how we simulate behavior from different models on these experiments, and (b) how we fit and compare these models to human data.

\subsection*{Scalable generation of cognitive tasks from LLMs}

Generating cognitive tasks from an LLM entailed a two-stage process. In the first stage, we query an LLM to synthesize the names for input features and targets. For instance, an example input feature in function learning could be \textsc{calorie intake} whose corresponding target is \textsc{weight}. In the second stage, the LLM is queried again but this time to generate numerical values for a given input feature and target pair generated from the first stage. That is, the LLM is tasked to generate different values for \textsc{[calorie intake, weight]}, for instance, \textsc{[2300, 152.0]} or \textsc{[1850,143.0]}. 
 
Below, we provide the prompts used in the two stages for the function learning domain; see SI for prompts for other domains. 
We used the following prompt to synthesize names for input features and targets for function learning: 
\begin{tcolorbox}[sharp corners, colback=mOrange!5!white,colframe=mOrange, width=0.95\textwidth, left=4pt,right=4pt, top=4pt, bottom=4pt, title=\textbf{Synthesize input feature name and its target} \label{prompt:task_label}]
I am a psychologist who wants to run a function learning experiment. In a function learning experiment, a real-world feature is mapped to its corresponding target, with both feature and target taking on continuous values.\\ 

Please generate names for features and its corresponding target for {\color{mOrange} \textbf{250}} different function learning experiments:\\

-- feature name, target name
\end{tcolorbox}

Next, we prompted the LLM to generate values for a function learning task generated from the first stage: 
\begin{tcolorbox}[sharp corners, colback=mBlue!5!white,colframe=mBlue, width=0.95\textwidth, left=4pt,right=4pt, top=4pt, bottom=4pt, title=\textbf{Generate values for a given function learning task} \label{prompt:task_values}]
I am a psychologist who wants to run a function learning experiment. For a function learning experiment, I need a list of features with their corresponding target. The feature in this case is {\color{mBlue} \textbf{calorie intake}}. The features take on only numerical values and must be continuous. The target, {\color{mBlue} \textbf{weight}}, should be predictable from the feature values and must also take on continuous values.\\

Please generate a list of {\color{mBlue} \textbf{20}} feature-target pairs sequentially using the following template for each row:\\

-- feature value, target value %\\
% Please do not produce any units and shuffle the order of items in the list.
\end{tcolorbox}

We generated a dataset containing around \num{10000} different function learning tasks with each task consisting of $20$ data points from \textsc{Claude-v2} \cite{Anthropic2023-hr}. The temperature parameter was set to one to induce diversity, and all other parameters were set to their default values.  We chose \textsc{Claude} as it can process up to \num{100,000} tokens, is instruction-tuned, cost-effective, and performed well out of the box on most of our preliminary analyses; see SI for information about the other LLMs we considered.  
% We provide the prompts for other two domains in the SI. 

To use and analyze the generated cognitive tasks, we parse all necessary quantities from the output text from the LLMs using regular expressions and stored them in numerical format in comma-separated-value (csv) files. These stored csv files are the datasets we use for further analysis. We expand on the parsing expressions, data-processing steps, and also provide a qualitative analysis of synthesized input feature and target names (see Figures S1, S3, S4, and S6) in the SI.

% For brevity, we have included the prompts and model details for only the function learning domain in the Methods. We provide full details for category learning and decision making, including other LLMs used and additional design choices, in the SI.  

\subsection*{Verifying the ecological validity of LLM-generated cognitive tasks}

To test the ecological validity of the generated cognitive tasks, we resort to two approaches. We either compare certain key statistics between LLM-generated tasks and a real world baseline, whenever we have access to a reasonable dataset, or compare it to real world statistics expected or predicted by prior work. We will discuss these tests for each domain individually below. \\

\noindent \textbf{Function learning.}

    We compared the data distributional properties of the LLM-generated function learning problems with 60 real-world regression tasks curated by Lichtenberg and colleagues \cite{lichtenberg2017simple}. We downsampled all tasks in the dataset to a single input dimension by applying univariate feature selection using the F-statistic for regression, as implemented in \textsc{scikit-learn} \cite{scikit-learn}, and included only tasks without missing values and with at least one valid input dimension in our analysis. Note that each dataset was split into separate tasks of 25 datapoints each, yielding a collection of regression problems with fixed size for analysis.

    For both real and LLM-generated function learning tasks, we estimated the relative frequency of different function classes within the dataset. We did this by first fitting models of different function classes to each LLM-generated task and then, assigning the function class with the best fitting model to the given task. 
    
    Specifically, we considered models from four well-studied function families, namely, linear, exponential, quadratic, and sinusoidal, from the literature \cite{busemeyer2013learning}; see SI for their exact model instantiations. The parameters of these models, $\phi$, were fit to data from the task to minimize the sum of squared errors (SSE) using the curve fit function from the \textsc{Scipy} optimization library \cite{2020SciPy-NMeth}. We then computed the Bayesian Information Criterion (BIC) for the fitted models from each function class, compared them against each other, and assigned the label of the function class that won the model comparison to a given task. Assuming $\hat{y}(\phi)$ and $y$ correspond to predicted target from the fitted model with parameters $\phi$ and true target, respectively, BIC computation entailed the following steps:

    \begin{equation}
    \begin{aligned}
    \text{SSE} &= \min_{\phi} \sum_{i=1}^{N} \left(y_i - \hat{y}_i(\phi)\right)^2 \\
    \text{BIC} &= N \cdot \ln(\text{SSE}) + |\phi| \cdot \ln(N)
    \end{aligned}
    \label{eqn:bic_mse} 
    \end{equation}
    where $|\phi|$ is number of parameters in a given model parameters, and $N$ is number of data points per task. This SSE-based approximation of the BIC assumes that model errors are Gaussian with constant variance, under which the negative log-likelihood is proportional to the sum of squared errors. 
    
    We obtained the proportion of different function classes by computing a histogram over the assigned class labels for all tasks in a given dataset. Furthermore, we assessed whether the fitted slope term of the linear model were predominantly positive and whether the fitted offset term, from the same model, was close to zero. \\

\noindent \textbf{Category learning.}
   We compared the data distributional properties of the LLM-generated category learning tasks with a real-world classification benchmark \cite{chan2022data}. For this, we used the OpenML-CC18 benchmarking suite, a curated collection of real-world classification tasks \cite{oml-benchmarking-suites}. We downsampled all tasks in the OpenML-CC18 benchmark to four feature dimensions by applying univariate feature selection using the ANOVA F-test implemented in \textsc{scikit-learn} \cite{scikit-learn} and included only binary classification tasks without any missing features in our analysis -- amounting to 28 tasks. 
   %In addition, we also contrasted LLM-generated tasks to synthetically-generated category learning tasks with linear and non-linear decision boundaries.

    We analyzed these collections of tasks in terms of their learning curves, input feature correlations, sparsity of predictive features, and linearity of the category structure. We obtained the learning curves by fitting a logistic regression model on a trial-by-trial fashion. For input correlations, we computed Pearson's correlation coefficient between every pair of features in the task. To get an estimate for task sparsity, we fitted a logistic regression model on the full data for each task and analyzed the sparsity of the resulting regression weights $\mathbf{w} \in \mathbb{R}^d$ using the Gini coefficient $G$: 
    \begin{equation}
        G(\mathbf{w}) = \dfrac{\displaystyle\sum\limits_{i=1}^d \displaystyle\sum\limits_{j=1}^d |\mathbf{w}_i - \mathbf{w}_j|}{2 d \displaystyle\sum\limits_{i=1}^d \mathbf{w}_i}
        \label{eqn:gini}
    \end{equation}
    
    For determining the linearity of the category structure, we fitted a logistic regression model and a logistic regression with second-order polynomial features on the full data $\mathcal{D}$ from each task. We then computed the BIC for both models and used them to approximate the posterior probability that the linear model offers a better explanation of the data (assuming a uniform prior over models), see Equation \ref{eqn:bic_linearity}. 

    \begin{align}
        p(M = {\text{linear}} | \mathcal{D}) \approx& \frac{\exp(-0.5 \cdot \text{BIC}_\text{linear})}{\sum_{m \in \{\text{linear, polynomial} \} } \exp(-0.5 \cdot \text{BIC}_m)}
        \label{eqn:bic_linearity}
    \end{align}

\noindent \textbf{Decision making.} We examined the distribution of input feature correlation, sparsity in predictive features, rank ordering of feature importance, and directionality of the features for the three LLM-generated decision making datasets belonging to ranking, direction, and unknown condition; see SI for details about their generation. For baseline, we considered the LLM-generated dataset in the unknown condition, as it allows contrasting dataset from the rank and direction condition with one that lacks explicit manipulation. See SI for data-distributional properties of LLM-generated decision making tasks.

For measuring correlation between input features, we compute pair-wise Pearson's correlation coefficient, following the same procedure we used in the domain of category learning; see Figure S7 (first column) in the SI for visualization. 

To measure sparsity of task features, we followed the same procedure as in the category learning task but instead of a logistic regression model, we fitted a linear regression model that predicts the continuous valued targets from the task features; see Figure S7 (second column) in the SI for visualization. 

For examining the rank ordering of feature importance, we fit a linear regression model, predicting the target from the input features. We then identified the feature with the highest absolute regression coefficient for each task and performed histogram over these positions to assess how often each feature was most predictive. If the intended manipulation was successful, we expect that the first feature should most frequently have the largest coefficient, followed by the second, and so on, reflecting a consistent ordering of feature relevance; see Figure S7 (third column) in the SI for visualization.
 
We assessed the directionality of each feature by examining the sign of the fitted regression coefficients from linear models as described above. If all coefficients are positive, it suggests that the intended manipulation was successful; see Figure S7 (fourth column) in the SI for visualization.

\subsection*{Ecologically Rational Meta-learned Inference}

Having generated and tested the ecological validity of LLM-generated cognitive tasks, we then trained transformer-based models on those tasks to derive explicit in-context learning models adapted to the ecological task distribution.
For this, we let a transformer-based model \cite{vaswani2017attention} auto-regressively predict a target, $y_t$ which can either be a discrete category or a continuous response, for a given input, $x_t$, conditioned on all preceding input-target pairs, $(x_{1: t-1}, y_{1:t-1})$. After the model predicts targets for all inputs in the sequence, the parameters of model, $\theta$, is updated based on the following objective:  
\begin{equation}
        \ell = \sum_t -\log p_{\theta} \left(y_{t} \mid x_{1: t}, y_{1:t-1}\right)
        \label{eq:loss}
\end{equation}
where $p_{\theta}$ defines the output probabilities produced by the model. 

The model is then trained until convergence, such that post convergence it can perform in-context learning. That is, the model can learn to predict the correct target for a new input based on previously seen input-target pairs -- provided in context. Critically, in-context learning is implemented by the model purely via its internal activations, without any additional weight updates after training.  
Previous work has demonstrated that this form of explicit in-context learning algorithms approximates the Bayes-optimal learning algorithm on the distribution of tasks $p(x_{1: T}, y_{1: T})$ encountered during training \cite{ortega2019meta}. This key result enables us to make links between in-context learning displayed by our models and rational analysis \cite{anderson1991human}. 
% \\

% \noindent \textbf{Model Architecture:} 
The base neural network in our in-context learning models was the transformer-based decoder architecture \cite{vaswani2017attention} with a causal attention mask, as done previously \cite{lake2023human, jagadish2024human, schubert2024context}. The network settings were chosen based on a hyper-parameter search and were different for each domain (see SI for details), but all models irrespective of domain used positional encoding based on sine and cosine functions of different frequencies \cite{vaswani2017attention}.  
% \\
% \noindent \textbf{Model Training:} 
For training, a batch of tasks is sampled from $p(x_{1: T}, y_{1: T})$ in each episode and the model predicts the target for the given input conditioned on all preceding input-target pairs. After which, model parameters are updated based on the objective mentioned in Equation \ref{eq:loss} using a schedule free optimizer \cite{defazio2024road} with the learning rate set to $3e-4$. We provide additional details about the model architecture and training procedure in the SI. 

\subsection*{Baseline models} We chose several domain-specific cognitive models and compared them with ERMI; see SI for full details. 
For function learning, we compared against a meta-learned inference (MI) model trained on functions drawn from a hand-crafted prior distribution over kernels. Following Lucas et al. \cite{lucas2015rational}, the prior probabilities for positive linear, negative linear, quadratic, and radial basis kernels were set proportional to 8, 1, 0.1, and 0.01, respectively.

We considered six models for the domain of category learning, namely, the rational model of categorization (RMC\cite{Anderson1991-ii}); a meta-learned inference (MI) model trained on synthetically generated problems with linear decision boundary; a meta-learned inference model trained on synthetically generated tasks with non-linear decision boundary (PFN\cite{muller2022transformers}); the generalized context model (GCM\cite{nosofsky1986attention}), a prototype model (PM\cite{homa1984role}), and a rule-based learning model (Rule\cite{mcdaniel2005conceptual}).

Four models were considered for the decision making task. 
% First, an ideal observer model \cite{binz2022heuristics}. 
First, a meta-learned inference model trained on synthetic decision making problems sampled from the true generative model used in the experiment (MI\cite{binz2022heuristics}). Second, a single-cue decision maker (SC\cite{binz2022heuristics}). Third, equal weighting decision maker (EW\cite{binz2022heuristics}). Fourth, a feedforward neural network (NN\cite{binz2022heuristics}).

\subsection*{Model simulations} In this section, we provide details of how model simulations were performed for the different experiments reported in this study. \\

\noindent \textbf{Function learning}

\noindent  \textit{Learning difficulty and speed.} To generate the learning difficulty curves shown in Figure \ref{fig:function_learning}C-D, we first sampled functions, $y$, from linear positive, linear negative, exponential, quadratic, and periodic families, for different values of input, $x$, ranging between 0 and 1. For the linear functions, we used the functional form $y=mx+c$, where we sampled slope and intercept terms from uniform distribution between -1 and 1. For the exponential functions, we used $y=a*e^{bx+c} + d$, where the terms were all sampled from uniform distribution between -1 and 1. Quadratic functions used the following parameterization: $y=w^2+c$, with values for parameters sampled from uniform distribution between -1 and 1. We used the functional form: $y=w*sin(2\pi x -\phi) + c$ for periodic function with amplitude, frequency, phase and offset sampled from uniform distribution between -1 and 1. All values were chosen such that final values are in the range between -1 to 1. We obtained the targets for each input value auto-regressively, conditioned on previous inputs and targets. We run this simulations 100,000 times for both ERMI and MI, and report the mean over trials for both models. 

\noindent \textit{Interpolation and Extrapolation.} We considered the same exact linear functions with fixed offset as used in the original Kwantes et al. \cite{kwantes2006extrapolation} study. We only additionally normalized the input and target to be between -1 and 1, such that it matches the range of inputs taken by ERMI during training. We extracted ERMI's and MI's predictions auto-regressively, conditioning on previously observed input-target values. \\

\noindent \textbf{Category learning} 

\noindent \textit{Learning difficulty.} To run simulations of the Shepard study \cite{Shepard1961-yu} on ERMI and MI, the geometric inputs used in the original study were converted into binary coded vectors taking values along the three input feature dimensions. The value assignment for a input feature was randomized in every run, the order of presentation of the input was also randomized, and the number of presentations of a input per block was matched to the original study. 
% In each run, the model was evaluated on a task of one particular type.

\noindent \textit{Learning strategy.}  The 616 choices made by ERMI and MI were divided into 11 blocks of 56 trials each. The choices were obtained from the model by simulating them on a numerically abstracted version of the task, similar to the learning difficulty mentioned above. The simulations were run for a total of 50 runs using the softmax temperature term fitted to participants in the Devraj et al. 2021 \cite{devraj2021dynamics} study. We then fit prototype-model (PM) and exemplar-based model (GCM) onto the choices of humans and models to see if they are better explained by prototype or exemplar-based strategy. To fit their parameters, we minimize the sum of squared errors (SSE) between observed and predicted probabilities for each participant for a given block following the original study's approach: 
    \begin{equation}
        SSE = \sum_{t=1}^{14} (p(y_t=1|x_t) - \hat{p}_{1,x_t})^2
    \end{equation}
where $p(y=1|x_t)$ is the predicted probability from the model -- either GCM or PM -- that input $x_t$ belongs to category 1 based on an entire trial segment (56 trials) of data, and $\hat{p}_{1,x_t}$ is the proportion of trials in the trial segment (out of those in which input $x_t$ was seen) in which the participant or model categorized input $x_t$ to category 1. We used SciPy's Sequential Least Squares Programming (SLSQP) method to obtain the best fitting parameter for the two models as in the original study \cite{devraj2021dynamics}. We then compared the SSE computed using the best-fitting parameters between the two models as shown in Figure \ref{fig:category_learning}.

\noindent \textit{Generalization.} We simulated ERMI and MI on the Johansen et al. study \cite{Johansen2002-xe}  for inverse temperature values, from zero to one in steps of 0.1, for a total of 544 runs. The models interacted with each of the nine training inputs 32 times, with the ordering of the inputs shuffled between runs. Predictions for the transfer inputs were derived by concatenating them -- one at a time -- at the end of 32 training blocks in every run. By doing so, we were able to derive the model's prediction for each unseen input around $77$ times.  In Figure \ref{fig:category_learning}, we reported average choice probabilities for the models using the inverse temperature value that minimized the pair-wise Euclidean distance between the human and model's choice probabilities. \\

\noindent \textbf{Decision making}

\noindent We evaluated ERMI and MI model on paired comparison tasks following the same generative model used as the original study \cite{binz2022heuristics}. ERMI and MI took values for the four attributes for both options along with the correct target from the previous trial as input at each step. They then predicted one of the two option on the current step. The simulation was performed for the same number of trials and blocks as in the original study. 
% The values for the attributes were normalized to be in the range xx and yy and target value for binary. 

% To infer the heuristic, which the models used, we used the Gini coefficient computed over the weights for the different attributes produced by the model as output; see Figure \ref{fig:decision_making} (A).

\subsection*{Model fitting and comparison}
Parameters for models considered in this work were fit to the data using maximum likelihood estimation. The exact model parameters fitted for each model and their implementation details are discussed in the SI.

After fitting the models, we performed a Bayesian model comparison, with goodness-of-fit to human choices measured based on posterior model frequency \cite{rigoux2014bayesian}. The posterior model frequency measures how often a given model offers the best explanation in the population. We computed it using a Python implementation of the Variational Bayesian Analysis (VBA) toolbox \cite{daunizeau2014vba}; see SI for additional details.  

\subsection*{Data, software, and code}
Data, code, and analysis scripts are available at \href{https://github.com/akjagadish/meta-learning-ecological-priors-from-llms/tree/main}{https://github.com/akjagadish/meta-learning-ecological-priors-from-llms/}
% Parts of the text were refined with the help of generative AI tools, such as ChatGPT and Writefull, which provided suggestions for rewording, paraphrasing, and restructuring. All outputs were carefully reviewed and edited before use.

\section*{Acknowledgements}
We thank the members of the Institute for Human-Centered Artificial Intelligence (HCAI) for their comments, discussions, and support throughout this work. Special thanks to Milena Rmus for helping out with Figure 1. 
We also specifically thank Devraj et al. \cite{devraj2021dynamics}, Nosofsky et al. \cite{Nosofsky1994-hw}, Binz et al. \cite{binz2022exploration}, Little et al. \cite{little2024function}, and Badham et al. \cite{Badham2017-hc} for making their data publicly available.   
This work was supported by the Max Planck Society, Helmholtz Center, Volkswagen Foundation, Princeton University, and Deutsche Forschungsgemeinschaft (DFG) under the German Excellence Strategy - EXC 2064 / 1 - 390727645. 
% Acknowledgements should be brief, and should not include thanks to anonymous referees and editors, or effusive comments. Grant or contribution numbers may be acknowledged.

\section*{Author contributions statement}
A.K.J., M.B., and E.S. conceived the study and developed the methodology and theoretical framework; J.C. and M.T. contributed to refining the theoretical framework; A.K.J. designed and conducted the experiments with contribution from M.B.; A.K.J. collected and preprocessed the data and generated the figures; A.K.J. wrote the original draft of the manuscript; A.K.J., J.C., M.T., E.S., and M.B. reviewed and edited the manuscript; E.S. acquired funding.

\clearpage
\newpage

\begin{center}
    \LARGE \textbf{Supplementary Information}\\[1em]
\end{center}
\setcounter{figure}{0}
\renewcommand{\figurename}{Figure}
\renewcommand{\thefigure}{S\arabic{figure}}
\renewcommand{\tablename}{Table}
\renewcommand{\thetable}{S\arabic{table}}
\section*{Function Learning}
\subsection*{LLM-generated tasks}
    The exact prompts and data generation pipeline for function learning are discussed in the Methods section of the main text. 
    \paragraph{Parsing synthesized task features and labels:}
        We queried \textsc{Claude-v2} to generate feature names in the format: \textsc{feature dimension 1, feature dimension 2, ...}; see Methods in the main text for exact prompts. To extract these, we used a series of regex patterns, such as \\ \verb|([A-Za-z&]+),([A-Za-z&]+)| and its higher-arity extensions, designed to capture up to five comma-separated alphanumeric feature names (including symbols like “\&”). These patterns allowed us to reliably extract structured feature descriptions across tasks. The parsed feature names were stored in a dataframe for subsequent task construction and evaluation.

\begin{figure*}[ht]
     \centering
     \includegraphics[width=1.\textwidth]{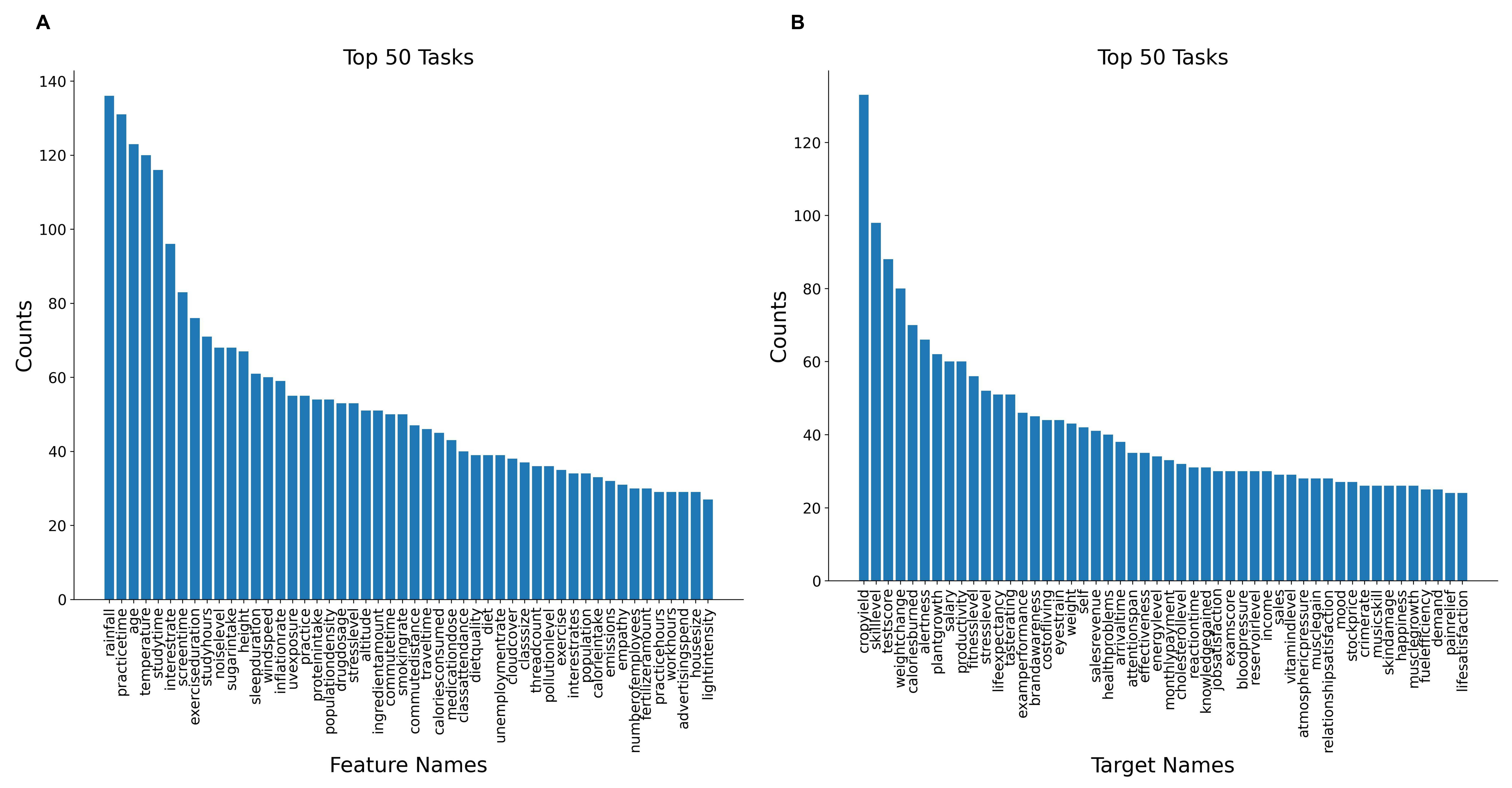}
      \caption{\textbf{Frequency of input and target labels in \textsc{claude-v2}  synthesized function learning tasks:} Counts for the top-50 most frequently occurring (a) input and (b) target labels computed over 9991 LLM-generated function learning tasks. These distributions confirm that the LLM-generated tasks capture real-world functional relationships.}
     \label{supp:functionlearning_features}
     \end{figure*}

    \paragraph{Qualitative analysis of synthesized task features and labels:}
    We show the counts for the top-50 most frequently occurring names for (a) inputs and (b) targets in Figure \ref{supp:functionlearning_features}. We found that LLM tends to produce input-target pairs belonging to everyday topics such as education (practice time versus skill), health (calories burned versus weight change), agriculture (rainfall versus crop yield), etc. 
     
    \paragraph{Parsing generated task data points:}
        \textsc{Claude-v2} was prompted to generate datapoints in the format: \textsc{- feature value 1, feature value 2, ..., feature value N, target value}; see Methods in the main text for exact prompts. To extract numerical values from these responses, we constructed regex expressions of the form \verb|([\d.]+),| repeated for each feature dimension, followed by \verb|([\d.]+)| for the target value. This pattern reliably captured sequences of decimal numbers across varying dimensionalities. The extracted values were stored in a dataframe, serving as a structured dataset for training and evaluating meta-learned function approximators.
    \paragraph{Data processing:}
        We filter out all tasks containing more than 20 data points to ensure consistent task lengths and evaluation settings. We randomly shuffled the trial order within each task.  We resampled the trials with replacement to match the target task duration, enabling evaluation on experiments with longer trial horizons without performance degradation. All feature dimensions were independently normalized to lie within $[-\text{scale}, \text{scale}]$ using a Min-Max normalization scheme, where $\text{scale} \in [0.1, 0.5]$ was fixed or randomly sampled. LLM-generated tasks can sometimes be of varying lengths, and in the case that the task length was shorter, they were padded with zeros to match the longest task in the batch. The maximum steps or number of trials for the experiment we considered was 25 trials. The batch size was fixed to 64 unless otherwise specified.
    \paragraph{Models fit to LLM-generated data:} We considered models from four well-studied function families, namely, linear, exponential, quadratic, and sinusoidal, as mentioned in the Methods. For the linear function, we chose the instantiation $y= a*x + b$, with initial parameters set to 1 and 0 for the slope and offset terms, respectively. For quadratic, we chose $y=a*x^2 + c$, with initial parameters for slope and offset set to 1 and 0 respectively. We chose $y= a*exp(b*x) + d$ for the exponential family, with initial parameters for $a$ set to the mean difference between the maximum and minimum of the target values, $b$ set to 1, and offset term set to the minimum of the target values. We chose $y=a*sin(b*x) + d$ for the sinusoidal family with initial parameters $a$ set to the mean difference between the maximum and minimum of the target values, $b$ set to $2*\pi$, and offset term set to the mean of the target values. We fit the parameters of these models to LLM-generated functions using the curve fit function from the \textsc{Scipy} optimization library \cite{2020SciPy-NMeth}. 

\subsection*{Human studies}

  \paragraph{Kwantes and Neal 2006 \cite{kwantes2006extrapolation}.}  In this study, 14 participants had to learn to predict values along the y-axis for different values on the x-axis, with samples drawn from a linear function $y = 2.2x + 30$. Before test phase, they were trained on 20 samples on the x-axis drawn such that their values on the y-axis were always in the range between 30 and 70. The samples were fixed but their order used for training was randomized per participant and session. In each trial, participants made their prediction by entering their estimate as numbers and locking in their answer by clicking on a button labeled ``submit your answer". After locking in, feedback was provided regarding their performance (in terms of accuracy score out of 100). Once training was complete, participants were shown 45 samples in the range from 0 to 100 and asked to enter their estimates. The presentation of the 45 samples were blocked into three sets of 15: low (0-30), medium (30-70), and high (70-100) range. The order in which the blocks were presented and the order of samples within them was randomized for each participant. 

  \paragraph{Little et al. 2024 \cite{little2024function}.} This study was conducted on 177 participants. The particular experiment we consider, called function estimate test, was included as part of larger paper-based questionnaire. In this experiment,  participants were presented 24 scatter plots, each depicting data from a different fictional scientific experiment, on a piece of paper, with two 7.5 cm by 7.5 cm graphs in each page with 4 cm gap between them. They were then instructed to draw the true underlying causal function for the data points in the graph. The data points could be presented in either large (zoomed in version), where the data points covered the entire figure, or small (zoomed out version), where it covered 40 percent of the total area, scale.  The relative position of the points in the small- and large-scale sets was kept identical. Three functions were used to generate data for the scatter plots, namely, linear, quadratic, or cubic polynomial functions. A small amount of Gaussian noise was added as jitter in all graphs. The data points and the drawn functions used for model fitting were extracted from scans of the physical document using a software program called Data Thief \cite{tummers2006datathief}. After extraction, the data was down-sampled to include 40 evenly spaced data points in the range of the x-axis and with all points scaled to be between -1 and 1. We used the data from the following \href{https://github.com/knowlabUnimelb/FUNCTION_ESTIMATION}{GitHub repository}. 
  
\subsection*{Hand-crafted tasks}
        \paragraph{Functional priors from rational model of function learning\cite{lucas2015rational} used for training MI model:} We generate 10,000 synthetic regression tasks for function learning using a mixture of kernels adapted from the study by Lucas et al. \cite{lucas2015rational}. Each task involved a one-dimensional input sampled from a uniform grid of 20 points in the interval $[-1, 1]$. The target output was computed by sampling a kernel type from a hand-crafted prior: favoring positive linear (probability 0.8), followed by negative linear (0.1), quadratic (0.01), and radial basis (0.001) functions and applying the corresponding transformation to the input. Parameters for each kernel (e.g. weights, intercepts, distances) were drawn from a gamma distribution with shape 1.001 and scale 1.0. A small amount of Gaussian noise was added to the target. All inputs and targets were dynamically scaled to lie in $[-\text{scale}, \text{scale}]$, where the scale is sampled from a uniform distribution in the range $[\text{0.1}, \text{0.5}]$ in each training batch.
\subsection*{Model architecture, and training}
    Each trial in a function learning task consisted of an input vector concatenated with the previous target value, and these were embedded into a 64-dimensional space. Positional encoding was applied using sine and cosine functions of varying frequencies, following Vaswani et al. \cite{vaswani2017attention}. A causal attention mask ensured that predictions at each time step were conditioned only on past inputs and targets. These masked sequences were processed using a Transformer decoder composed of six layers, with 64-dimensional embeddings, eight attention heads, and 256 hidden units in the feedforward layers. The decoder outputs were passed through two independent linear projections to produce the mean and standard deviation of a normal distribution. The negative log-likelihood (NLL) was computed over all targets in a given batch, and minimizing it served as a loss function for training the network parameters. The model parameters were updated using the \textsc{ScheduleFree} optimizer \cite{defazio2024road} with a baseline learning rate of $3 \times 10^{-4}$. Each model was trained for \num{250,000} episodes, with periodic evaluation on held-out tasks to monitor generalization performance.
    
\subsection*{Model fitting and comparison}
   We did not fit any model parameters to human data in both ERMI and MI. We computed the response from these models by querying it on new inputs while being conditioned on the input-target pairs participant observed before drawing the functions. For model comparison, we report the mean-squared error between the participant's actual response, sampled from the functions they drew through the data points displayed to them, and model predicted target for the same input.  

\subsection*{Additional results}

\begin{figure*}[!h]
    \centering
    \includegraphics[width=\textwidth]{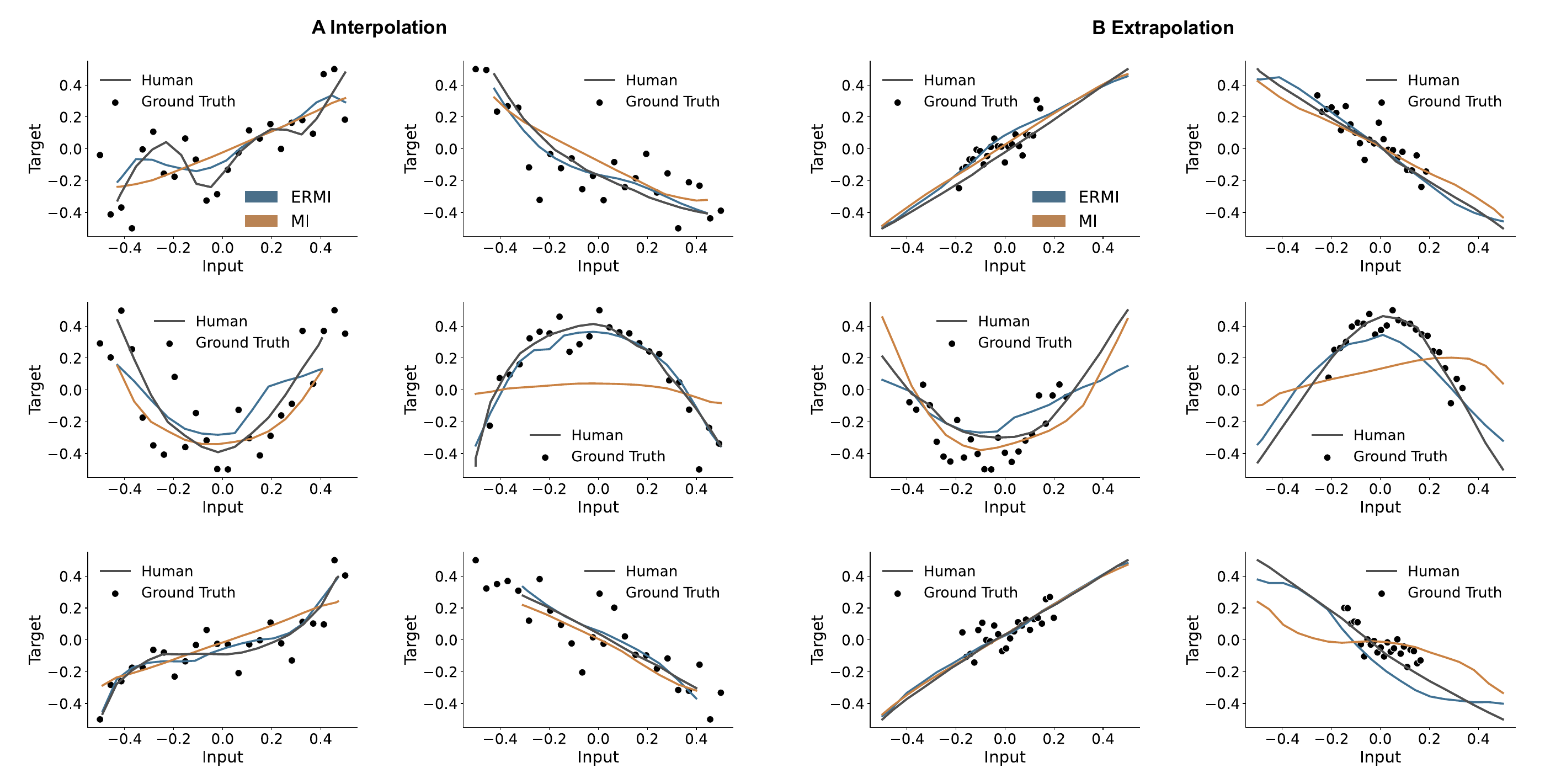}
    \caption{\textbf{Predictions derived from ERMI and MI for function families used in the Little et al. study}\cite{little2024function} for both interpolation (zoomed in; A) and extrapolation (zoomed out; B) condition. The function families considered were linear (top row), quadratic (middle row) and cubic polynomial (bottom row); see Human studies section in Methods for details.}
    \label{fig:functionlearning}
    % \vspace{-0.20cm}
    \end{figure*}
    
\clearpage
\newpage
\section*{Category Learning}
\subsection*{LLM-generated tasks}
    \paragraph{Prompts:} We used the following prompt to synthesize feature names and category labels for the category learning task.  
    \begin{tcolorbox}[sharp corners, colback=mBlue!5!white,colframe=mOrange, width=0.95\textwidth, left=4pt,right=4pt, top=4pt, bottom=4pt, title=\textbf{Synthesize feature names and category labels} \label{prompt:task_label_categorylearning}]
    I am a psychologist who wants to run a category learning experiment. In a category learning experiment, there are many different three-dimensional stimuli, each of which belongs to one of two possible real-world categories.\\ 

    Please generate names for three stimulus feature dimensions and two corresponding categories for 250 different category learning experiments:
    \end{tcolorbox}
     
    %% Step 2: generating the task features 
    In the second stage, we prompted the LLM to generate data points for the synthesized features and the category label. Below is the prompt, for a category learning where the synthesized input features were sodium, fat, and protein, and categories are healthy or unhealthy:
    
    \begin{tcolorbox}[sharp corners, colback=mBlue!5!white,colframe=mBlue, width=0.95\textwidth, left=4pt,right=4pt, top=4pt, bottom=4pt, title=\textbf{Generate category learning tasks} \label{prompt:task_values_categorylearning}]
        I am a psychologist who wants to run a category learning experiment. For a category learning experiment, I need a list of stimuli and their category labels. Each stimulus is characterized by three distinct features: {\color{mBlue}\textbf{sodium}}, {\color{mBlue}\textbf{fat}}, and {\color{mBlue}\textbf{protein}}. These features can take only numerical values. The category label can take the values {\color{mBlue}\textbf{healthy}} or {\color{mBlue}\textbf{unhealthy}} and should be predictable from the feature values of the stimulus.\\
    
        Please generate a list of 100 stimuli with their feature values and their corresponding category labels using the following template for each row:\\
        
        -- feature value 1, feature value 2, feature value 3, \\ \phantom{-} category label 
        % \textit{-- feature value 1, 2, 3, } \textit{category label}
        \end{tcolorbox}
    \paragraph{Parsing synthesized task features and labels:}
    We prompted \textsc{Claude-v2} to generate task features and labels in the format: \textsc{feature dimension 1, feature dimension 2, ..., feature dimension N, category label 1, category label 2}. We extracted relevant entries using the regex pattern \verb|\d+.(.+?)\n|, which captures text following a numbered bullet point up to the first newline. The resulting string was split at the commas to separate feature names from category labels. All parsed information was stored in a dataframe for downstream use.

\begin{figure*}[!h]
     \centering
     \includegraphics[width=\textwidth]{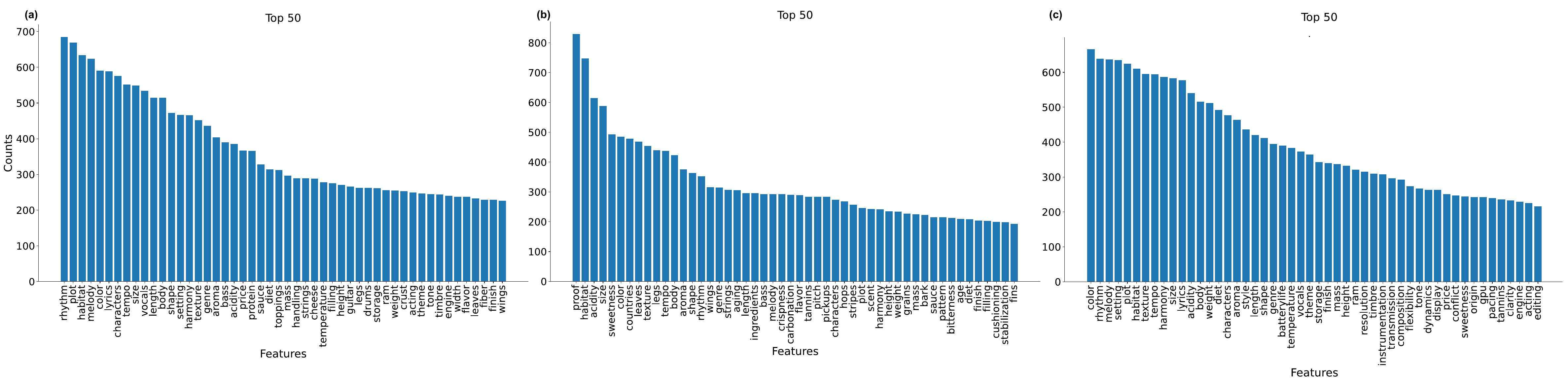}
      \caption{\textbf{Frequency of input feature names in \textsc{claude-v2} synthesized category learning tasks:} Counts for the top-50 most frequently occurring input features in the 23421, 20690, and 13693 category learning tasks generated for three (a), four (b), and six-dimensional features respectively. These distributions confirm that the LLM-generated real-world features relevant for category learning.}
     \label{supp:LLMfeatures}
     \end{figure*}

\begin{figure*}[!h]
     \centering
     \includegraphics[width=\textwidth]{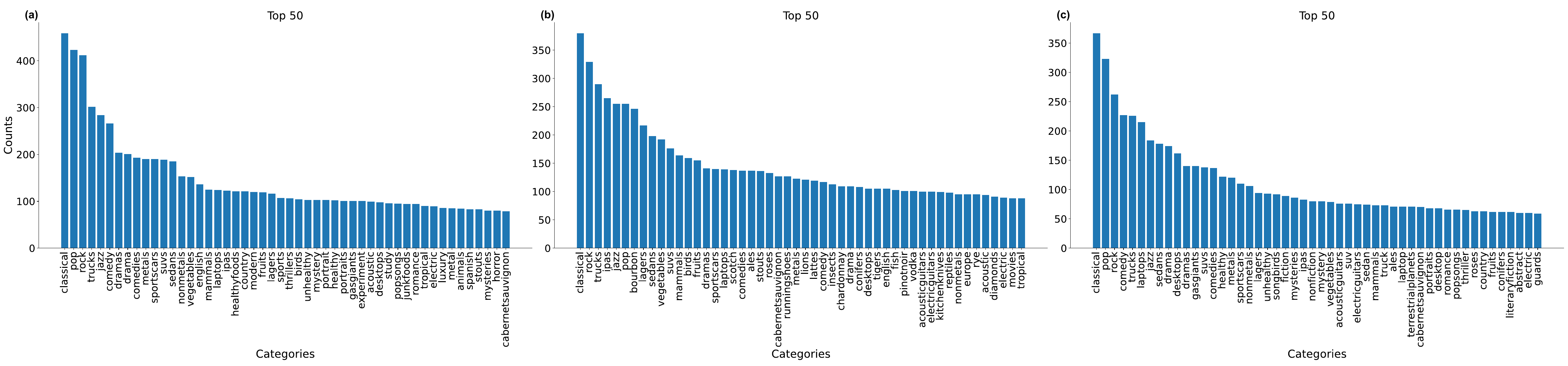}
      \caption{\textbf{Frequency of different category names in  \textsc{claude-v2}  synthesized category learning tasks:}  Counts for the top-50 most frequently occurring category names in the 23421, 20690, and 13693 category learning tasks generated for three (a), four (b), and six-dimensional features respectively. These distributions confirm that the LLM-generated category names prevalent in the real-world.}
     \label{supp:LLMcategories}
     \end{figure*}

    \paragraph{Qualitative analysis of synthesized task features and labels:}
    We show the counts for the top-50 most frequently occurring input feature names in Figure \ref{supp:LLMfeatures} and category names in Figure \ref{supp:LLMcategories} for the 23421, 20690, and 13693 category learning tasks generated with three (a), four (b) and six-dimensional input features, respectively. When it comes to input feature names, we found that the LLM tends to produce features belonging to topics such as musicality (like rhythm, melody, lyrics, tempo, vocals), food (like aroma, texture, crust, diet, protein), etc.  With regard to category names, there were also many related to music (for example, classical, pop, jazz, rock), but also vehicles (like trucks, SUVs, sedans), technology (laptops, desktops, iPads), etc. 
     
    \paragraph{Parsing generated task data points:}
    To generate data points for each task, we queried \textsc{Claude-v2} using the format: \textsc{- feature value 1, feature value 2, ..., feature value N, category label}. The model reliably followed this format. To parse the resulting output, we used a suite of regex patterns designed to handle diverse data formats, including numeric values (with or without decimals), alphanumeric labels, hyphens, and various delimiters. Table \ref{regex-table} lists all the regex patterns employed. These enabled us to successfully parse 95\% of the generated tasks. The parsed values were stored in a dataframe, forming an offline task repository to train the ecologically rational meta-learned inference model.
    
    \paragraph{Data pre-processing:} We filter out all tasks with more than two unique category labels and then binarize the category labels, which are originally strings, to make them consistent across tasks. The assignment of category labels, that is either `0' or `1', within a category learning task was randomized during batch creation. This ensures that there can be no unintended correlations between the inputs seen during training and the labels (across all training data each input vector is assigned half of the time to label ‘0’ and half of the time to label ‘1’). We also normalized each feature independently using a min-max normalization scheme such that values taken by any feature lie always between zero and one. Both the task features and data points were shuffled while generating tasks. Note that the tasks generated by LLMs are typically of different lengths.  Whenever the sampled tasks are of variable lengths, they are padded with zeros to match the length of the longest task sample within the batch. We additionally also sampled LLM-generated data points with replacement to match the length of the experimental task used in the Devraj et al. 2021 \cite{devraj2021dynamics} and Johansen et al. 2002 \cite{Johansen2002-xe} studies. We resorted to this strategy as the LLM-generated tasks had a maximum of about 200 data points per task and by resampling, we can evaluate the model on experiments with larger horizons without any drop in performance. The batch size was set to 64 for three- and four-dimensional inputs and to 32 for six-dimensional inputs and it operated under a maximum steps regime of 400, 300, and 650 for three, four, and six-dimensional tasks respectively. 

    \begin{table}[!h]
    \caption{Regular expression patterns used for parsing the data points generated for category learning tasks by \textsc{claude-v2}}
    \label{regex-table}
    \vskip 0.15in
    \begin{center}
    \begin{small}
    \begin{sc}
    \begin{tabular}{ll}
    \toprule
    Index & Regular expression \\
    \midrule
    1 & \verb|([\d.]+),([\d.]+),([\d.]+),([\w]+)| \\
    2 & \verb|([\w\-]+),([\w\-]+),([\w\-]+),([\w]+)| \\
    3 & \verb|([-\w\d,.]+),([-\w\d,.]+),([-\w\d,.]+),([-\w\d,.]+)| \\
    4 & \verb|([^,]+),([^,]+),([^,]+),([^,]+)| \\
    5 & \verb|([^,\n]+),([^,\n]+),([^,\n]+),([^,\n]+)| \\
    6 & \verb|(?:.*?:)?([^,-]+),([^,-]+),([^,-]+),([^,-]+)| \\
    7 & \verb|([^,-]+),([^,-]+),([^,-]+),([^,-]+)| \\
    8 & \verb|r'^(\d+):([\d.]+),([\d.]+),([\d.]+),([\d.]+),([\w]+)'| \\
    9 & \verb|r'^(\d+):([\w\-]+),([\w\-]+),([\w\-]+),([\w\-]+),([\w]+)'| \\
    10 & \verb|r'^(\d+):([-\w\d,.]+),([-\w\d,.]+),([-\w\d,.]+),([-\w\d,.]+),([-\w\d,.]+)'| \\
    11 & \verb|r'^(\d+):([^,]+),([^,]+),([^,]+),([^,]+),([^,]+)'| \\
    12 & \verb|r'^(\d+):([^,\n]+),([^,\n]+),([^,\n]+),([^,\n]+),([^,\n]+)'| \\
    13 & \verb|r'^(\d+):(?:.*?:)?([^,-]+),([^,-]+),([^,-]+),([^,-]+),([^,-]+)'| \\
    14 & \verb|r'^(\d+):([^,-]+),([^,-]+),([^,-]+),([^,-]+),([^,-]+)'| \\
    15 & \verb|^(\d+):([\d.]+),([\d.]+),([\d.]+),([\d.]+),([\d.]+),([\d.]+),([\w]+)|\\
    16 & \verb|^(\d+):([\w-]+),([\w-]+),([\w-]+),([\w-]+),([\w-]+),([\w-]+),([\w]+)|\\
    % 10 & \verb|(\d+):([-\w\d,.]+),([-\w\d,.]+),([-\w\d,.]+),([-\w\d,.]+),([-\w\d,.]+),([-9 & 11 & \w\d,.]+),([-\w\d,.]+)'|\\
    17 & \verb|(\d+):([^,]+),([^,]+),([^,]+),([^,]+),([^,]+),([^,]+),([^,]+)|\\
    18 & \verb|(\d+):([^,\n]+),([^,\n]+),([^,\n]+),([^,\n]+),([^,\n]+),([^,\n]+),([^,\n]+)|\\
    19 & \verb|(\d+):(?:.*?:)?([^,-]+),([^,-]+),([^,-]+),([^,-]+),([^,-]+),([^,-]+),([^,-]+)|\\
    20 & \verb|(\d+):([^,-]+),([^,-]+),([^,-]+),([^,-]+),([^,-]+),([^,-]+),([^,-]+)|\\

    \bottomrule
    \end{tabular}
    \end{sc}
    \end{small}
    \end{center}
    \vskip -0.1in
    \end{table}

\subsection*{Human studies}
  \paragraph{Nosofsky et al. 1994 \cite{Nosofsky1994-hw}.} In their replication of the Shepard et al. 1961 \cite{Shepard1961-yu} study, Nosofsky and colleagues conducted the study on 120 participants. The authors used geometric inputs that varied in shape (squares or triangles), interior line type (solid or dotted), and size (large or small). In total, 40 participants performed each of the six category structures, considered in Shepard et al. 1961 \cite{Shepard1961-yu}. The participants were informed that the rules for each problem were independent. Following the same methodology as Shepard et al., the learning process involved classifying inputs into two categories and receiving feedback. This process was repeated over several blocks (containing up to 16 trials) with randomized input order in each block. Learning in the task was measured until participants achieved a no-error streak in four consecutive sub-blocks of eight trials or reached a maximum of 400 trials. In tasks belonging to \textsc{type 1}, inputs were assigned to a category depending on the values they take along one of the three dimensions, whereas in \textsc{type 2} tasks, inputs were assigned to a category by applying the exclusive-or rule along two relevant dimensions. Category assignment in tasks belonging to \textsc{type 3}, \textsc{type 4}, and \textsc{type 5} used a unidimensional rule-plus-exception structure with some inputs grouped in the central region and some in the periphery. Lastly, \textsc{type 6} tasks required considering feature values along all dimensions, and they require the memorization of every item and its associated category to solve them correctly. For the illustration of category structures for the six types, please refer to Figure 1 in Nosofsky et al. study \cite{Nosofsky1994-hw}. \\
  
  \paragraph{Badham et al. 2017 \cite{Badham2017-hc}.} In this study, the authors partially replicated the original Shepard et al. 1961 \cite{Shepard1961-yu} study by running it on 96 adults aged between 18 to 87 years. As inputs, they used eight geometric shapes varying in size (large or small), shape (square or triangle), and color (black or white) shown on a mid-gray background. The order of inputs and their category assignment were randomized. Unlike the original study, the authors only considered the first four types of category structures but with the advantage that all participants performed all four types. Participants performed each task type for a total of six blocks with each block containing 16 trials (resulting in a total of 96 trials) or until they reached a criterion of perfect performance in two consecutive blocks.  \\

   \paragraph{Smith et al. 1998 \cite{smith1998prototypes}.} The study was run on 32 participants, where each participant was presented 14 different six-dimensional inputs, with each input mapping to a six-letter nonsensical word such as gafuzi, kafitdo, nivety, wysero, etc (see Appendix A of \cite{smith1998prototypes} for all words). For modeling, we represented each input as a six-digit binary string, where each digit and position corresponds to a specific letter. For instance, assuming the input `gafuzi' corresponds to the binary code `000000', `gyfuzi' corresponds to `010000', and so on. The inputs were assigned to categories such that input `000000' corresponds to category 1 and input `11111' corresponds to category 2. 
   %Note that "prototypes were created randomly but with several constraints to ensure the pronounceability of all inputs, the orthographic appropriateness of all inputs, the identical syllabification of all inputs, and the roughly equal use of all vowels" \cite{smith1998prototypes}. 
    In this work, we only considered data from the non-linearly separable (NLS) category structure from Experiment 2. In this category structure, each category consisted of six inputs with five features in common with the prototype, and one input with five features in common with the opposing prototype. For instance, if category 1 contained seven inputs as follows: [000000, 100000, 010000, 001000, 000010, 000001, 111101]. The remaining seven inputs belonged to category 2 [111111, 011111, 101111, 110111, 111011, 111110, 000100]. Participants had to categorize a input into one of these two categories and had unlimited time to make their choices. After making their choice, they were told if it was a correct decision or not. Participants completed a total of 560 trials over 10 blocks of 56 trials each. In each block, participants saw each input four times.  \\

    \paragraph{Devraj et al. 2021 \cite{devraj2021dynamics}.} Devraj and colleagues replicated a study of Smith et al. 1998 \cite{smith1998prototypes} and collected data from 60 participants. Participants were recruited from the 18-23 age range and English-speaking population using Prolific. Their study involved 11 blocks and had 616 trials in total. We used the data from the following \href{https://github.com/arjundevraj/rational-categorization}{GitHub repository}.  \\
    % We used data from this study as data on the level of trial-level choices is online available, which is not the case for the original study. 

     \paragraph{Johansen et al. 2002 \cite{Johansen2002-xe}.} Johansen and colleagues conducted their categorization study on 130 participants in which they presented four-dimensional inputs with each dimension taking binary values. Each of the inputs was a computer-generated drawing of a rocket that varied along four binary-valued dimensions: The shape of the wing (triangular or rectangular), tail (jagged or boxed), nose (staircase or half-circle), and porthole (circular or star) \cite{Johansen2002-xe}. The authors used the same category structure as those used in previous studies  \cite{Medin1978-xf, Nosofsky1994-gu}. This category structure is ill-defined in that no single feature along a dimension can be used to perfectly classify inputs. 
     Instead, the categories have a family resemblance structure in that category 1 inputs tend to have a value of 0 along each dimension, and category 2 inputs tend to have a value of 1 along each dimension. More concretely, they assigned they five inputs [0001, 0101, 0100, 0010, 1000] to category 1 and the remaining four inputs [0011, 1001, 1110, 1111] to category 2. The inputs were presented serially with their order randomized within each block.  Participants had unlimited time to make their choice and  were informed whether or not it was a correct choice after each choice. Participants completed a total of 288 training trials, or 32 blocks of 9 trials each, in which they saw each input once. In addition to the training block, participants had to perform a transfer block after 2, 4, 8, 16, 24, and 32 blocks of training. In a transfer block, the eight training inputs along with eight other unseen transfer inputs were shown without corrective feedback. The encoding for transfer inputs, labeled T1 to T7, were (in order): [1011, 1010, 0111, 1101, 1100, 0110, 0000]. It is the category assigned in the transfer block, which is of major interest in this work. \\

\subsection*{Hand-crafted tasks}

    \paragraph{Bayesian logistic regression prior used for training MI model:}
    We generated \num{10,000} synthetic binary classification tasks with a linear decision boundary using a Bayesian logistic regression model. To do this, we sample the input features from a normal distribution with zero mean and unit variance for a given number of data points and input dimensions. We then applied a linear transformation, followed by a sigmoid function, and rounded the result to determine the binary class for the given input. The parameters of the linear transformation are sampled from a normal distribution with zero mean and unit variance. The maximum number of data points within a task was set to 400, 650, or 300 for category learning tasks with three-, four-, and six-dimensional inputs, respectively. These values were chosen according to the length of the experiments on which these models were evaluated.

    \paragraph{Bayesian neural network prior used for training prior-fitted network (PFN) model:}
    We generated \num{10,000} synthetic binary classification tasks using a version of the Bayesian neural network (BNN) developed by Müller et al. \cite{Muller2021-ol}. We used normally-distributed i.i.d. input features for a given number of data points and input dimensions. We then passed the input through a BNN with two layers with tanh non-linearity and hidden dimensionality of 64. The network weights and biases were sampled from a normal distribution with a mean of zero and a standard deviation of 0.1 and subjected to an additional sparsity constraint (i.e., 20 percent of randomly chosen network weights and biases set to zero). The maximum number of data points was once again set to 400, 650, or 300 for category learning tasks with three-, four-, and six-dimensional inputs, respectively. The model output is passed through a sigmoid function to generate probability estimates, which are then rounded to determine the class for the given input. 
    
\subsection*{Model architecture, and training}
    % Steps involved after passing the model to transformer decoder
     The task features, which contain values for the different input features and the target from the previous trial, were mapped to a 64-dimensional embedding space and positional encoded using sine and cosine functions of different frequencies as in  Vaswani et al. \cite{vaswani2017attention}. Then a causal attention mask was generated for the inputs so that the model makes conditional predictions on all preceding data points. The inputs along the attention mask are then passed to the transformer decoder model, which has six layers, a model dimension of 64, 256 hidden units in the feed-forward network, and eight attention heads. The output of the transformer was then passed through a linear readout and sigmoid function to generate probability estimates for category 1. In practice, inference for all time steps is performed in parallel by passing a causal attention mask to the transformer decoder module in \textsc{PyTorch} \cite{NEURIPS2019_9015}. We used binary cross-entropy (BCE) loss for a given batch of inputs and updated the model parameters using the \textsc{ADAM} optimizer \cite{kingma2014adam} with a learning rate of $10^{-4}$. We trained all our models for a total of \num{500,000} episodes. 

\subsection*{Baseline models}

    Apart from models derived by meta-learning on hand-crafted priors, we considered four other cognitive models as baselines in the domain of category learning, as detailed below. 

    \paragraph{Rational model of categorization (RMC):} 
    The RMC is a Bayesian model of human category learning developed by Anderson et al. \cite{Anderson1991-ii}. To derive this model, we simulated data from underlying generative model, such that it followed the data-generating distribution described in Badham et al. \cite{Badham2017-hc}, and meta-learned on the generated data, similar to meta-learning on hand-crafted priors. The architecture and training of the model followed the protocol used for ERMI, MI and PFN. We set the free parameters for the RMC based on an earlier study \cite{Nosofsky1994-hw} to the following values: $c=0.318$, $s_P=0.488$, and $s_L=0.046$. However, we did not account for these parameters in our model comparisons, which could explain why the predictive performance RMC is overestimated. 
    
    \paragraph{Prototype-based model (PM):} Over the years, many different versions of the prototype model have been produced \cite{Medin1978-xf, smith1998prototypes}. We used the version from Smith et al.  \cite{smith1998prototypes}.  This model assigns a category to an observed stimulus based on the similarity distance to the prototype from each category. Specifically, the similarity distance between the stimulus and a prototype, $q_k$, for category $k$ is calculated as a weighted sum of absolute differences in the dimensions of the features $n$, with $w_j \in [0, 1]$ corresponding to the weights per feature. The weights are normalized to sum up to 1 as shown in Equation \ref{eq:distance}.
    \begin{equation}
        d_{x, q_k}=\sum_{j=1}^n w_j\left|x_j-q_{k, j}\right|,
        \label{eq:distance}
    \end{equation}
    The prototypes themselves can be learned or directly specified during model definition. In our case, we assume the prototypes for the two categories $\{q_1, q_2\}$ as a learnable parameter and learn them during the model fitting procedure. That is,  $q_{k,j} \in [0, 1.]~ \forall j = \{1, 2,...n\}$ are assumed to be learnable model parameters.
    The similarity distance between prototypes and stimuli is converted into a psychological space using:
    \begin{equation}
    \eta_{x,q_k} = e^{-c \cdot d_{x,q_k}}
    \label{eq:similarity}
    \end{equation}
    where $c$ is a sensitivity parameter that can shrink or amplify discriminability in a psychological space. The probability of the stimulus being assigned to the category $k=1$ was then calculated using the following.
    \begin{equation}
        P(k=1 \mid x) = \frac{\eta_{x,q_1}}{\eta_{x,q_1} + \eta_{x,q_2}}
        \label{eq:prob_assignment}
    \end{equation}

    Furthermore, the predicted likelihood of the final model is a mixture between the predicted probability of the model and a random guess, with the guessing parameter $\epsilon$ controlling the mixture probabilities.
    \begin{equation}
        p(k=1 \mid x) = (1 - \epsilon) P(k=1 \mid x) + \epsilon \cdot \mathrm{K}^{-1}
        \label{eq:gcm_pm}
    \end{equation}
    where $\mathrm{K}$ indicates the number of categories.
    
    \paragraph{Generalized context model (GCM):} GCM is an exemplar-based model of human category learning developed by Nosofsky et al. \cite{nosofsky1986attention}. The GCM assigns an observed stimulus to a category by comparing the sum of its similarity scores to all previously seen exemplars in each category, $\{C_1, C_2\}$. The raw distance between the observed stimulus and the exemplars and the similarity score were calculated based on Equations \ref{eq:distance} and \ref{eq:similarity}, respectively. The posterior probability of category membership $k=1$ is calculated on the basis of normalized similarity scores as follows.
    \begin{equation}
        P(k=1 \mid x) = \frac{\sum_{y \in C_1} \eta_{x,y}}{\sum_{y \in C_1} \eta_{x,y} + \sum_{y \in C_2} \eta_{x,y}}
    \end{equation}

    The final likelihood of category membership is computed as a mixture between the estimate of posterior probability and a random guessing model as mentioned in Equation \ref{eq:gcm_pm}.

     \paragraph{Rule:} The rule model considered as the baseline in this work assigns a stimulus to a category based on one of the two rules, whichever better explains the choices of the participants. The first rule is based on the values taken by stimulus features along one dimension, and the second is based on the application of the conjunctive rule on pairs of features, whether a given pair of stimulus features takes on the same value. The final category membership is determined by a mixture between the predicted posterior class probabilities of the model and a random guess, as discussed in Equation \ref{eq:gcm_pm}.
    
    % \paragraph{Rule plus exception model (Rulex):} For the rulex model, we used the same implementation as the rule model but provided exceptions to the rules following the design of the human experiment as input to the model. Specifically, for the Devaraj et al. \cite{devraj2021dynamics} task, we provide $[1, 1, 1, 1, 0, 1]$ as the exception stimulus for category 1, and for category 2, it was set to $[0, 0, 0, 1, 0, 0]$. For the Badham et al. \cite{Badham2017-hc} task, we provide $[1, 1, 1]$ and $[0, 0, 0]$ as exceptions to the \textsc{type-2} task. The final category probabilities predicted by the model were a mixture between the predicted likelihoods of the model and a random guessing model, as shown in Equation \ref{eq:gcm_pm}. The version of Rulex model used in this work is a simplified version of the one from Nosofsky et al. \cite{Nosofsky1994-gu} where the model learns the exceptions along with the rule. 

\subsection*{Model fitting}
   The parameters of all models in the domain of category learning were fit to human data using maximum likelihood estimation. We explain the exact implementation details for the different model classes in the following.  The complete list of the parameters fitted for each model is shown in Table \ref{table:model_params}. 
    
    \paragraph{MI, PFN, RMC and ERMI:}
    For models derived using meta-learning, we fitted the inverse temperature term $\beta$ within the sigmoid function, which squashes the output from the final layer of the transformer to be within $[0,1]$, to each participant. This term was set to a value of 1 during meta-learning to allow us to derive a Bayes-optimal model and was only fitted during the evaluation phase (bounded to be within $[0, 10]$), with the rest of the model weights frozen. For parameter fitting, we used the differential evolution optimizer available in the \textsc{SciPy} optimization library \cite{2020SciPy-NMeth}. 

    \paragraph{GCM and PM:} We fit the three parameters common to GCM and PM, namely feature weights, sensitivity, and the random guessing parameter, with feature weights bounded to lie within the range $[0,1]$ and summing to 1; sensitivity term bounded to lie within the $[0, 20]$ range; and the guessing parameter bounded to be within $[0,1]$. The prototype model also required learning the prototypical stimulus for each category, which is of the same dimensionality as the input stimulus, with the feature values bounded within $[0,1]$. For parameter fitting, we used the \textsc{minimize} module available in the \textsc{SciPy} optimization library. 
    
    \paragraph{Rule:} We used the same procedure as above except that we learn the stimulus dimension $v_i$ on which the rule is applied. 

    \paragraph{\textsc{Claude-v2}:} We used the same procedure as above except that only the guessing parameter, $\epsilon$, is learned. 
    
    \begin{table}[!h]
    \caption{This table provides the complete list of model parameters that were fit to human data in the domain of category learning, where $\beta$ is the inverse temperature term, $w_{i}$ indicates the weights for the stimulus feature dimension $i$, $n$ is the number of stimulus feature dimensions, $c$ is the sensitivity term, $\epsilon$ is noise term in an epsilon greedy policy, $q_1$ and $q_2$ are the values for the prototypes for $d$ stimulus features, and $v_i$ are the stimulus dimension on which the rule is applied.}
    \label{table:model_params}
    \vskip 0.15in
    \begin{center}
    \begin{small}
    \begin{sc}
    \begin{tabular}{lll}
    \toprule
    \textbf{Model}     & \textbf{Parameters} &\\  
    \midrule
    $\text{ERMI, MI, PFN, RMC}$ &  $\beta$ \\ 
    $\text{GCM}$ &   $c, \epsilon, w_{i}$ & $\forall ~i \in\{1,2, \ldots, n\}$ \\
    $\text{PM}$ &  $c, \epsilon, w_{i}, q_{1,i}, q_{2,i}$ & $\forall ~i \in\{1,2, \ldots, n\}$ \\
    % w_{1}, w_{2},...,w_{d}, g_{1}, g_{2},..., g_d, f_{1}, f_{2},..., f_d, $ \\ 
    $\text{Rule}$ &  $v_1, v_2, \epsilon $ \\ 
    % $\text{Rulex}$ &  $v_1, v_2, \epsilon$ \\ 
    $\text{\textsc{Claude-v2}}$ &  $\epsilon$ \\ 
    
    \bottomrule
    \end{tabular}
    \end{sc}
    \end{small}
    \end{center}
    \vskip -0.1in
    \end{table}

\subsection*{Bayesian model comparison}
    
   After fitting the model parameters to human data using maximum likelihood estimation, we computed the Bayesian information criterion (BIC), which penalizes model fitting performance based on its complexity, for models $m$ for a given participant as follows: 
    \begin{equation}
        \text{BIC}_{m} = -2 \cdot \max _{\theta_m} \sum_{t=1}^{T} \log p_{\theta_m} \left(\hat{y_{t}} \mid x_{1: t}, y_{1:t-1} \right) + |\theta_m| \log (T)
    \end{equation}

    where $|\theta_m|$ is the number of parameters estimated for the model $m$, $T$ is the number of trials in the task and $\hat{y_{t}}$ is the choice made by the participant in a given trial $t$. 
    
    Once computed, we compared the goodness-of-fit between models using posterior model frequency, which measures how often a given model offers the best explanation in the population. For computing it, we used a Python implementation of the Variational Bayesian Analysis (VBA) toolbox \cite{daunizeau2014vba}. The toolbox required providing log-evidences for each model and participant pair, which we approximate using $-0.5 \cdot \text{BIC}_{m}$; see Rigoux et al. study\cite{rigoux2014bayesian} for details about this model comparison procedure.

 \begin{table}[!h]
    \centering
    \caption{Mean performance of humans and models for each rule type in replication of \cite{Shepard1961-yu} study over 15 blocks. Human data was taken from Table 1 in \cite{Nosofsky1994-hw}.} 
    \vspace{2mm}
    \begin{tabular}{rccccccc}
    \toprule
     \multicolumn{1}{c}{\textbf{Model}} & \multicolumn{6}{c}{\textbf{Rule}}  & \textbf{MSE} \\
     \cmidrule(lr){2-7} 
      \multicolumn{1}{r}{}& \textit{Type 1} & \textit{Type 2}  &  \textit{Type 3} &  \textit{Type 4} &  \textit{Type 5} &  \textit{Type 6}  \\
    \hline
    Humans & .0201 & .0565  & .1015 & .1120 & .1212 & .2048 & .0000 \\
    \hline
    ERMI & .0586 & .0891 & .0855 & .0826 & .0888 & .1172 & \textbf{.0287 } \\
    \hline
    MI & .0686 & .4089 & .2404 & .1431 & .2880 & .4201 & .2627 \\
    \hline
    PFN & .0170  & .3405  & .1533  & .0226  & .2371  & .3975 & .1736  \\
    \hline
    RMC & .1329  & .2215  & .1903  & .1718  & .2132  & .3364 & .1003  \\
    \bottomrule
    \end{tabular}
    \label{table:shepardtask}
    \end{table}

% \subsection*{Additional Results}

 \subsection*{\textsc{Claude-v2} as a cognitive model of human category learning}

    To simulate the study by Badham et al. \cite{Badham2017-hc} using \textsc{Claude-v2}, we queried the model with the prompt shown below. Geometric stimuli from the original experiment were described in text format. The order of presentation of the stimulus was randomized and the number of presentations per block was compared to the original study. As the Claude API returns only sampled tokens, not log-probabilities, we coded predictions as binary outcomes, $\pi(k=1 \mid x_t; x_{1:t-1}, y_{1:t-1})$. The final model predicted category probabilities is again a mixture between the category prediction from the model and a random guess as mentioned in Equation \ref{eq:llm}. We conducted 96 simulation runs for each of the six categorization rules.
    \begin{equation}
        p(k=1 \mid x_t) = (1 - \epsilon) \pi(k=1 \mid x_t; x_{1:t-1}, y_{1:t-1}) + \epsilon \cdot \mathrm{K}^{-1}
        \label{eq:llm}
    \end{equation}

    \begin{tcolorbox}[sharp corners, colback=mBlue!5!white,colframe=mBlue!75!black, width=1.\textwidth, left=4pt,right=4pt, top=4pt, bottom=4pt, title=\textbf{Prompt for  Badham et al. 2017 study} \label{prompt:shepard1961task}]
       In this experiment, you will be shown examples of geometric objects. Each object has three different features: size, color, and shape. Your job is to learn a rule based on the object features that allows you to tell whether each example belongs in the \{A\} or \{B\} category. As you are shown each example, you will be asked to make a category judgment and then you will receive feedback. At first you will have to guess, but you will gain experience as you go along. Try your best to gain mastery of the \{A\} and \{B\} categories. \\
    
        - In trial 1, you picked category \{A\} for Big Black Square and category \{A\} was correct.\\
        - In trial 2, you picked category \{A\} for Small Black Triangle and category \{B\} was correct\\
    
    Human: What category would a Small Black Triangle belong to? (Give the answer in the form ``Category $\langle$your answer$\rangle$").
    
    Assistant: Category
    \end{tcolorbox}

    \begin{figure*}[!h]
    \centering
    \includegraphics[width=\textwidth]{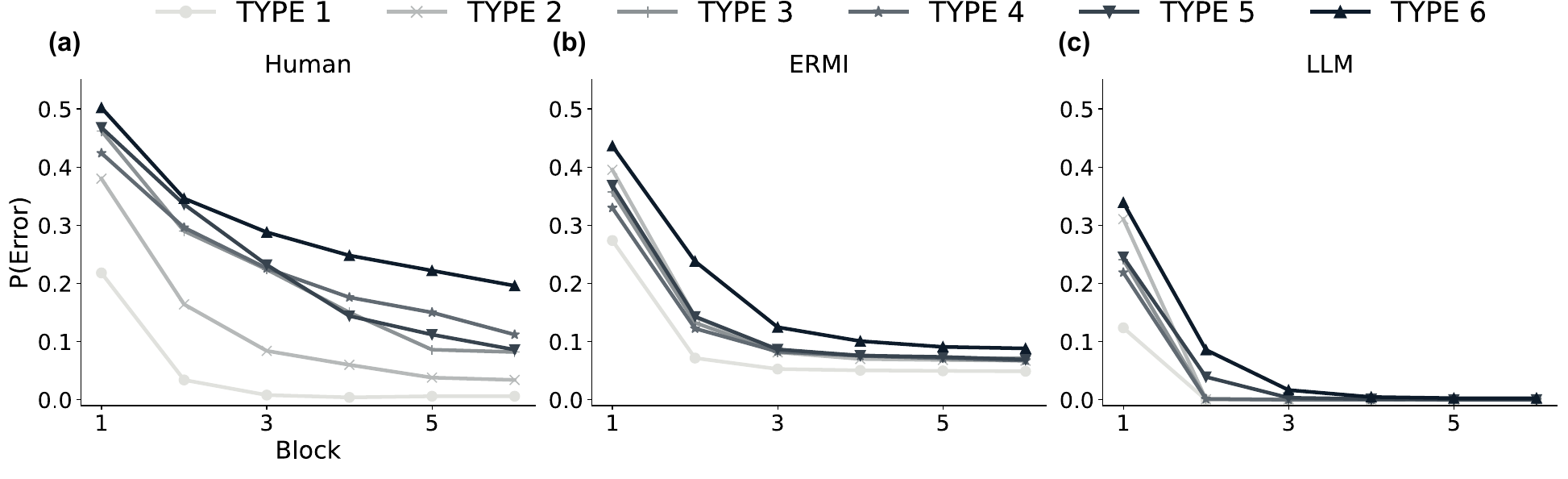}
    \caption{\textbf{Unlike ERMI, \textsc{Claude-v2} does not show human-like learning difficulties}: \textbf{(a-c)} Average error probabilities for each task \textit{type} in each block of 16 trials for (a) humans, (b) ERMI, and (c) LLM. Human data in (a) was reproduced from Table 1 in Nosofsky et al. \cite{Nosofsky1994-hw} study. ERMI was simulated on \textit{type 1-6} tasks for 50 runs with the inverse temperature set to $\beta=0.4$. \textsc{Claude-v2} was simulated for 94 runs each on \textit{type 1-6} tasks with temperature term set to 0.}
    \label{fig:llmsimulationsbadham2017}
    \vspace{-0.20cm}
    \end{figure*}

\clearpage
\newpage
\section*{Decision Making}
\subsection*{LLM-generated tasks}
    \paragraph{Prompts:}  In the following, we provide the prompts used in the two stages of decision making learning domain. 
    We used the following prompt to synthesize the names of stimulus features and targets, similar to function learning, separately for each of the three conditions, ranking, direction, and unknown. 
    
    \begin{tcolorbox}[sharp corners, colback=mOrange!5!white,colframe=mOrange, width=0.95\textwidth, left=4pt,right=4pt, top=4pt, bottom=4pt, title=\textbf{Synthesize stimulus feature name and its target for ranking condition} \label{prompt:task_label_ranked}]
    I am a psychologist who wants to run a function learning experiment. In a function learning experiment, a real-world feature is mapped to its corresponding target, with both feature and target taking on continuous values.\\
    
    Please generate names for features and its corresponding target for {\color{mOrange}\textbf{250}} different function learning experiments. Additionally, order the feature names according to how well they predict the target:\\
    
    -- feature name, target name
    \end{tcolorbox}
    
    \begin{tcolorbox}[sharp corners, colback=mOrange!5!white,colframe=mOrange, width=0.95\textwidth, left=4pt,right=4pt, top=4pt, bottom=4pt, title=\textbf{Synthesize stimulus feature name and its target for direction condition} \label{prompt:task_label_direction}]
    I am a psychologist who wants to run a function learning experiment. In a function learning experiment, a real-world feature is mapped to its corresponding target, with both feature and target taking on continuous values.\\
    
    Please generate names for features and its corresponding target for {\color{mOrange}\textbf{250}} different function learning experiments. Additionally, the features should be such that higher feature values lead to higher target values:\\
    
    -- feature name, target name
    \end{tcolorbox}
    
    \begin{tcolorbox}[sharp corners, colback=mOrange!5!white,colframe=mOrange, width=0.95\textwidth, left=4pt,right=4pt, top=4pt, bottom=4pt, title=\textbf{Synthesize stimulus feature name and its target for unknown condition} \label{prompt:task_label_v0}]
    I am a psychologist who wants to run a function learning experiment. In a function learning experiment, a real-world feature is mapped to its corresponding target, with both feature and target taking on continuous values.\\
    
    Please generate names for features and its corresponding target for {\color{mOrange}\textbf{250}} different function learning experiments:\\
    
    -- feature name, target name
    \end{tcolorbox}
    
    Next, we prompted the LLM to generate values for tasks generated from stage 1: 
    \begin{tcolorbox}[sharp corners, colback=mBlue!5!white,colframe=mBlue, width=0.95\textwidth, left=4pt,right=4pt, top=4pt, bottom=4pt, title=\textbf{Generate values for ranking condition} \label{prompt:task_values_ranking}]
    I am a psychologist who wants to run a function learning experiment. For a function learning experiment, I need a list of features with their corresponding target. The features in this case are feature1, feature2, feature3, and feature4. These features take on only numerical values and must be continuous. The target, <target>, should be predictable from the feature values and must also have continuous values. Note that the features are listed according to how well each of them can predict the target. The first feature is most useful for predicting the target, the second feature is the second most useful, etc.
    
    Please generate a list of <num-data> feature-target pairs sequentially using the following template for each row:
    - feature value 1, feature value 2, feature value 3, feature value 4, target value
    % Please do not produce any units and shuffle the order of items in the list.
    \end{tcolorbox}
    \begin{tcolorbox}[sharp corners, colback=mBlue!5!white,colframe=mBlue, width=0.95\textwidth, left=4pt,right=4pt, top=4pt, bottom=4pt, title=\textbf{Generate values for direction condition} \label{prompt:task_values_dir}]
    I am a psychologist who wants to run a function learning experiment. For a function learning experiment, I need a list of features with their corresponding target. The features in this case are feature1, feature2, feature3, and feature4. These features take on only numerical values and must be continuous. The target, <target>, should be predictable from the feature values and must also have continuous values. Note that the values taken by the features should be such that higher feature values lead to higher target values.
    
    Please generate a list of <num-data> feature-target pairs sequentially using the following template for each row:
    - feature value 1, feature value 2, feature value 3, feature value 4, target value
    % Please do not produce any units and shuffle the order of items in the list.
    \end{tcolorbox}
    \begin{tcolorbox}[sharp corners, colback=mBlue!5!white,colframe=mBlue, width=0.95\textwidth, left=4pt,right=4pt, top=4pt, bottom=4pt, title=\textbf{Generate values for unknown condition} \label{prompt:task_values_ranking_unk}]
    I am a psychologist who wants to run a function learning experiment. For a function learning experiment, I need a list of features with their corresponding target. The features in this case are feature1, feature2, feature3, and feature4. These features take on only numerical values and must be continuous. The target, <target>, should be predictable from the feature values and must also have continuous values.
    
    Please generate a list of <num-data> feature-target pairs sequentially using the following template for each row:
    - feature value 1, feature value 2, feature value 3, feature value 4, target value
    
    % Please do not produce any units and shuffle the order of items in the list.
    \end{tcolorbox}

    \paragraph{Parsing and pre-processing:} The parsing expressions used and the data preprocessing steps are the same as in the function learning domain.

\begin{figure*}[!h]
     \centering
     \includegraphics[width=1.\textwidth]{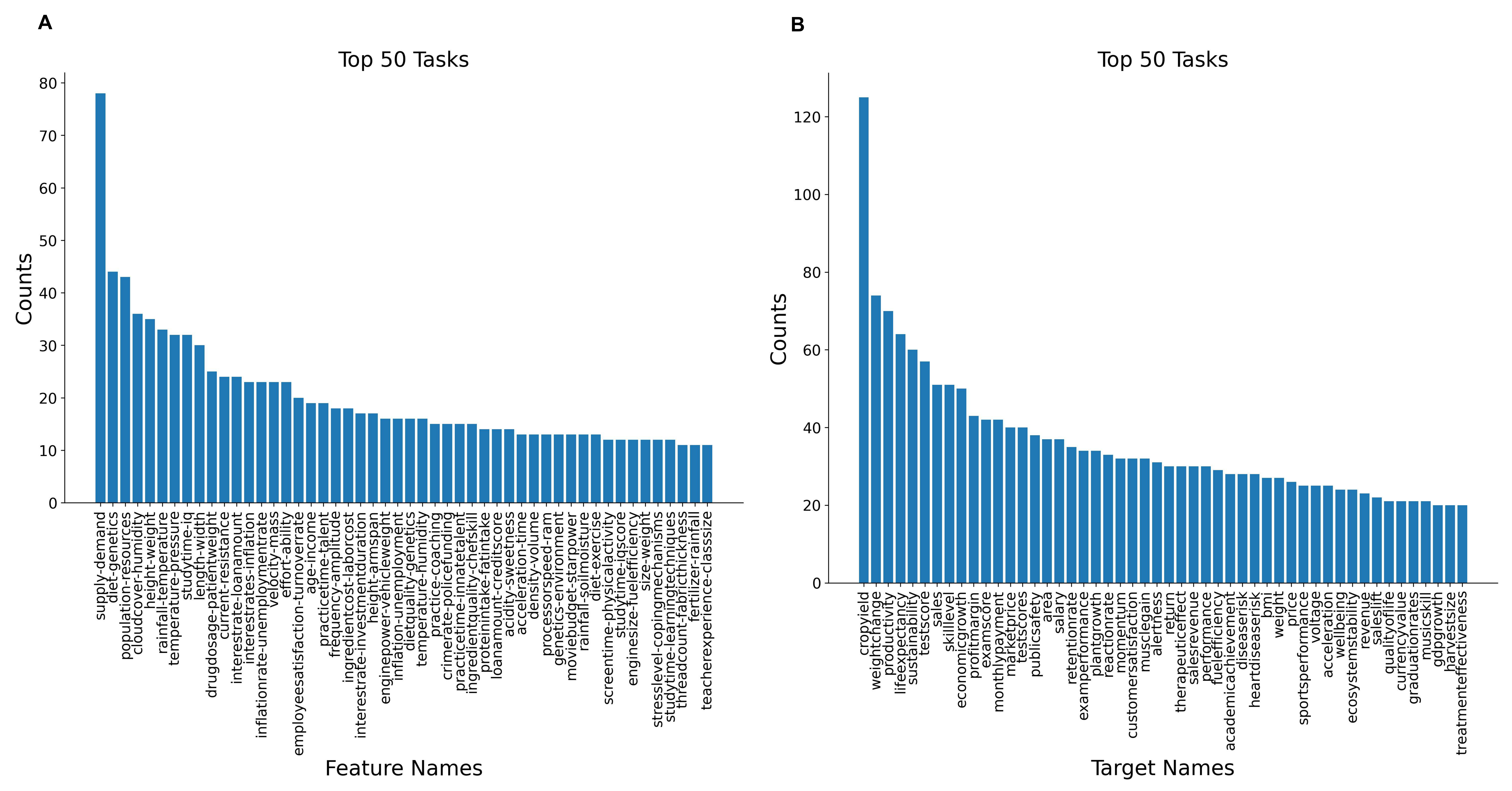}
      \caption{\textbf{Frequency of input and target labels in \textsc{claude-v2} synthesized decision making tasks:} Counts for the top-50 most frequently occurring (a) two-dimensional input feature names and (b) target names computed over 9254 LLM-generated decision learning tasks belonging to the unknown condition. These distributions confirm that the LLM-generates real-world functional relationships that are useful for everyday decision making.}
     \label{supp:decisionmaking_features}
     \end{figure*}

    \paragraph{Qualitative analysis of synthesized input features and labels:}
    We show the counts for the top-50 most frequently occurring names for (a) input features and (b) targets in Figure \ref{supp:decisionmaking_features}. We found that the LLM tends to produce input-target pairs that are relevant to everyday life such as supply-demand influence on productivity, diet-genetics influence on weight change, cloud cover-humidity on crop yield, study time-intelligence quotient on test score, etc.   
     
\begin{figure*}[!h]
    \centering
    \includegraphics[width=\textwidth]{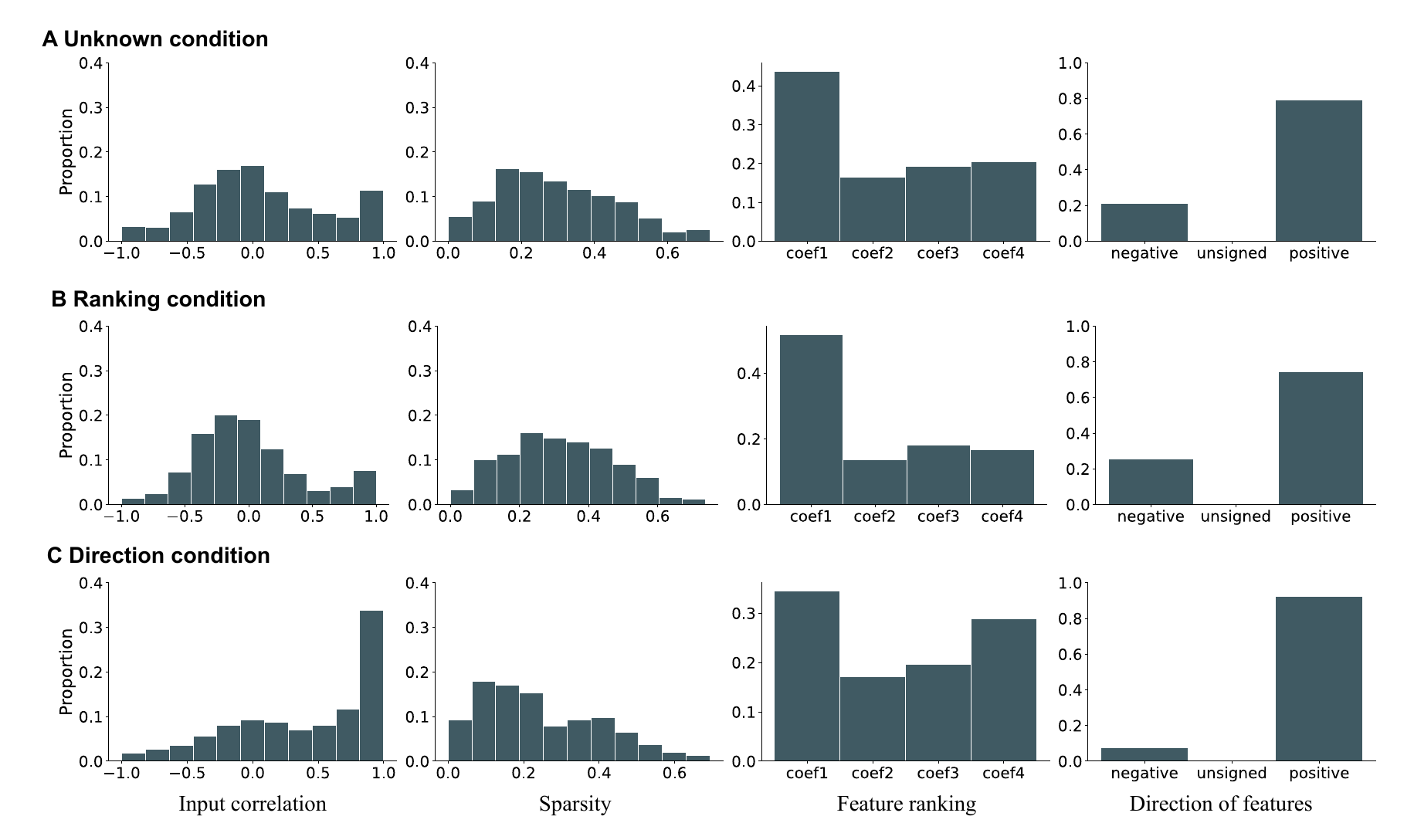}
    \caption{\textbf{Data-distributional properties of LLM-generated decision making tasks} for (A) unknown, (B) ranking and (C) direction condition. Histogram of Pearson correlation coefficients between all distinct pairs of normalized input features (first column). Histogram of Gini coefficients computed on the absolute ordinary least squares (OLS) weights when regressing the normalized target on all normalized inputs with an intercept (second column; higher values indicate sparser weights). Histogram of the index of the input feature with the largest absolute per-feature OLS weight, where each per-feature model regresses the target on a single feature with an intercept (third column; feature ranking). Histogram of the sign of per-feature OLS weights from those single-feature-with-intercept regressions (fourth column; direction). 
    % Tasks from these distributions were used to train ERMI for unknown, ranking, and direction condition respectively.
    }
    \label{fig:decisionmaking}
    \end{figure*}

\newpage
 \paragraph{Data-distributional properties of LLM-generated tasks:} We generate three datasets of decision-making tasks, one for each of (A) unknown, (B) ranking, and (C) direction, following the prompts described above. To examine their properties and verify if the manipulation was successful, we computed four key statistics: input correlations, sparsity in predictive features, ranking of feature importance, and directionality of features with respect to the target, and compared them across datasets. Specifically, we contrasted the ranking and direction conditions with the unknown condition, which served as a baseline. We found that the first feature was more often the most important feature in terms of predictive power (see the caption of Figure \ref{fig:decisionmaking} for details on the calculation) in the ranking condition ($51.76\%$) than in the unknown condition ($43.75\%$). Likewise, the proportion of features positively correlated with the target was higher in the direction condition ($92.46\%$) than in the unknown condition ($79\%$). 

\subsection*{Human studies}
     \paragraph{Binz et al. 2022 \cite{binz2022heuristics}.} This study was conducted on 27 participants in total, with each participant performing 30 different paired comparison tasks. Tasks were generated by first sampling feature weights from a standard normal distribution. Feature vectors for each option were then drawn from a multivariate normal distribution with zero mean and fixed covariance. Finally, the binary choice outcome was determined by sampling from a Bernoulli distribution, where the success probability was given by a probit regression over the difference in feature values (see Equation 2 in the main paper). The feature weights were kept the same within a task, which consisted of 10 trials, but were resampled between tasks. All participants performed the same set of paired comparison tasks but presented in randomized order. In Experiment 3a, participants observed two features per option, whereas in Experiment 3b they observed four features per option. In neither of these two experiments, information about the ranking of the features and their directions were provided. The experiment itself was framed as an alien sports competition on an unknown planet.  Participants observed two or four numerical attributes for two aliens, depending on the experiment they were part of. They indicated their choice by pressing a button corresponding to the alien they believed would most likely win. This cover story was used so that the meaning of the feature attributes remained abstract for each participant.  Participants were not told about the underlying feature weights, and they had to learn them through trial and error, using the feedback about correct choice provided after each trial.  All participants in the experiment performed a short tutorial and went through a comprehension check, which ensured clear understanding of the experimental protocol before data-collection. 
\subsection*{Hand-crafted tasks}
\paragraph{Synthetic paired-comparison tasks used for training MI:}
    We generated three synthetic datasets of paired-comparison problems (between 7000-9000 tasks per set) under \emph{ranking}, \emph{direction} and \emph{unknown} conditions. For each task, a weight vector $w \in \mathbb{R}^d$ was sampled from a standard normal distribution. In the \emph{direction} condition, weights were constrained to be non-negative by taking absolute values; in the \emph{ranking} condition, feature importance was rank-ordered by sorting weights by magnitude; and in the \emph{unknown} condition, weights were left unconstrained. To generate options, the feature vectors were sampled from a zero-mean multivariate normal distribution with covariance $\Sigma = L \, \mathrm{diag}(\theta) \, L^\top$, where $L$ was drawn from an LKJ (Lewandowski–Kurowicka–Joe distribution; $\eta = 2$) prior and $\theta = \mathbf{1}$. The LKJ distribution is a flexible prior over correlation matrices that allows control over the strength of correlations while ensuring positive definiteness. Each trial presented a pair of options $x_a, x_b \sim \mathcal{N}(0, \Sigma)$, with the comparison input defined as $x = x_a - x_b$. We randomly determine which option has the highest criterion by sampling from a Bernoulli distribution as follows: $y \sim \mathrm{Bernoulli}\!\left(\Phi(w^\top x/\sigma)\right)$ with $\sigma = 0.1$. Each task contained a maximum of 10 trials, which corresponded to the length of the experiment in which this model was evaluated.

\subsection*{Model architecture, and training}
    The input vector for a given trial in a decision making task was the difference between the input features for the two options, computed for each dimension independently, and the correct target option from the previous trial. The number of features in the decision making task was either two or four dimensions and the total number of observations in a given task was 20. These inputs were embedded into a 64-dimensional space, with positional encoding applied using sine and cosine functions of varying frequencies, following Vaswani et al. \cite{vaswani2017attention}. 
    A causal attention mask ensured that predictions at each time step were conditioned only on all previous inputs. These masked sequences were processed using a Transformer decoder composed of six layers, with 64-dimensional embeddings, eight attention heads, and 256 hidden units in the feedforward layers. The decoder outputs were passed through a linear projection to produce weights for the different feature dimensions. The likelihood of a target option is then calculated by first projecting the output through a linear layer, multiplying it element-wise with the current input features, summing across dimensions, and passing the result through a sigmoid to obtain a Bernoulli probability. Training was performed using the negative log-likelihood (NLL) loss over all input observations in a batch. The model parameters were updated as mentioned before using the \textsc{ScheduleFree} optimizer \cite{defazio2024road} with a baseline learning rate of $3 \times 10^{-4}$. Each model was trained for \num{100000} episodes, with periodic evaluation on held-out tasks to monitor generalization performance.

\subsection*{Baseline models}
 Apart from the MI model derived by meta-learning on tasks generated with hand-crafted priors, we considered three other cognitive models as baselines in the domain of decision making, as detailed below. 

 \paragraph{Single-cue decision maker (SC):} In Equation \ref{eq:single_cue}, we demonstrate formally how the heuristic of single-cue decision making makes a decision given the input feature. Note that $x^{*}$ indicates that the model only takes into account a single feature, which in this case was the most predictive feature. This means that only one parameter is fitted to human choices. 

 \begin{equation}
p(y_t = 1 \mid x_t, \theta_m, m=\text{SC}) 
= \Phi\!\left( \frac{\theta_m \cdot x_t^{*}}{\sqrt{2}\sigma} \right)
\label{eq:single_cue}
 \end{equation}

 where $\Phi$ is the cumulative distribution function of a standard normal distribution, $\theta_m$ is the weight of the selected feature, and $\sigma$ is the noise standard deviation. 

 \paragraph{Equal weighting decision maker (EW):} We considered a probabilistic version of the equal weighting model, as shown in Equation \ref{eq:equal_weighting}. When $w>0$, this model probabilistically selects the option with the larger sum of features. In contrast, when $w<0$, it selects the option with the smaller sum of features. Once again, only one parameter is fitted to the human data. 
 
  \begin{equation}
p(y_t = 1 \mid x_t, \theta_m, m=\text{SC}) 
= \Phi\!\left( \frac{\theta_m \cdot \sum_{i=1}^{d} x_{t,i}}{\sqrt{2}\sigma} \right)
\label{eq:equal_weighting}
 \end{equation}

  where $\Phi$ is the cumulative distribution function of a standard normal distribution, $\theta_m$ is the feature weight, and $\sigma$ is the noise standard deviation. 
    
 \paragraph{Feedforward neural network (NN):} We used a feedforward neural network from the Binz et al. \cite{binz2022heuristics} study as an additional baseline model. This model predicts the target given the difference between the input features of the two options and the previous target as input. The network consisted of a single hidden layer with 128 units followed by two linear transformations projected to the mean and (log) standard deviation of a normal distribution.  The neural network parameters were trained by gradient descent on the negative log-likelihoods of the target. During model fitting, the learning rate parameter and the inverse temperature term were fit to human choices; see Appendix F in Binz et al. \cite{binz2022heuristics} for implementation details. 

 \subsection*{Model fitting and comparison}

For fitting the model parameters, we performed the maximum likelihood estimation using Bayesian optimization \cite{gpyopt2016}, following the procedure used by Binz and colleagues. \cite{binz2022heuristics}.  A complete list of model parameters that are fitted to human choices can be found in Table \ref{table:dm_model_params}. Upon fitting, we followed the same exact steps as described above for category learning for Bayesian model comparison. That is, we used a VBA tool box, where we provide $-0.5 \cdot \text{BIC}_{m}$ as an approximation of log-evidence for each model and participant; see Rigoux et al. study\cite{rigoux2014bayesian} for details.

\begin{table}[!h]
    \caption{This table provides the complete list of model parameters that were fit to human data in the domain of category learning, where $\beta$ is the inverse temperature term, $\theta$ indicates the weights for the stimulus feature dimension, and $\alpha$ is learning rate.}
    \label{table:dm_model_params}
    % \vskip 0.15in
    \begin{center}
    \begin{small}
    \begin{sc}
    \begin{tabular}{lll}
    \toprule
    \textbf{Model} & \textbf{Parameters}\\  
    \midrule
    $\text{ERMI, MI}$ &  $\beta$ \\ 
    $\text{SC}$ &  $\theta$ \\ 
    $\text{EQ}$ &  $\theta$ \\ 
    $\text{NN}$ &  $\alpha, \beta$ \\ 
    
    \bottomrule
    \end{tabular}
    \end{sc}
    \end{small}
    \end{center}
    % \vskip -0.1in
    \end{table}

 \section*{Alternative LLMs}
 
 During the early stages of this work, we also considered two other LLMs: Llama-2 \cite{touvron2023llama} and GPT-4 \cite{achiam2023gpt}, which were among the best performing models at the time. However, the non-instruction-tuned Llama-2 (the only version available at the time) could not consistently produce the 100+ data points required for each category learning task. Its outputs were also difficult to parse, as they frequently failed to follow the specified format. More recently, with Llama-3.1 (70B) \cite{meta2024llama3}, we were able to generate decision-making datasets whose quality matched those produced by \textsc{Claude-v2}.  

Preliminary analysis with GPT-4 revealed that it often sampled input features from a uniform distribution, relying on its internal coding module. It also tended to generate only simple heuristic rules, such as requiring the sum of two features to exceed the third, or the mean of two features to be greater than another, for assigning an input to its category. Furthermore, statistical analysis on a small GPT-4 generated dataset showed that its task statistics closely resembled those of category learning tasks with hand-crafted priors (specifically Bayesian logistic regression prior). Due to this lack of diversity in the generated task statistics, we decided to use \textsc{Claude-v2} over GPT-4.

\bibliography{main}

% \authordeclaration{The authors declare no competing interests. 

\end{document}